\documentclass[pra,onecolumn,preprint,11pt,showpacs,citeautoscript,footinbib,eqsecnum,superscriptaddress]{revtex4-2}

\newcommand{\vect}[1]{\boldsymbol{#1}}

\usepackage[T1]{fontenc}
\usepackage[utf8]{inputenc}
\usepackage{lipsum}
\usepackage{amsmath}
\usepackage{amssymb}
\usepackage{bbm}
\usepackage{braket}
\usepackage{xcolor}
\usepackage{pifont}
\usepackage[mathscr]{euscript}
\usepackage[shortlabels]{enumitem}
\usepackage[justification=justified]{subcaption}
\usepackage{ragged2e}
\DeclareCaptionJustification{justified}{\justifying}
\usepackage{graphicx}
\usepackage{lipsum}
\allowdisplaybreaks
\usepackage{float}
\usepackage{graphicx}
\usepackage{dsfont}
\usepackage{comment}
\usepackage[colorlinks=true]{hyperref}  
\hypersetup{
    bookmarks=true,         
    unicode=false,          
    pdftoolbar=true,        
    pdfmenubar=true,        
    pdffitwindow=false,     
    pdfstartview={FitH},    
    pdftitle={Flstar},    
    pdfauthor={Tikhanovskaya, Sachdev, Samajdar},     
    pdfsubject={},   
    pdfcreator={},   
    pdfproducer={}, 
    pdfkeywords={} {} {}, 
    pdfnewwindow=true,      
    colorlinks=true,       
    linkcolor=blue, 
    citecolor=blue,        
    filecolor=magenta,      
    urlcolor=blue           
} 
\usepackage[papersize={8.5in,11in}]{geometry}

\allowdisplaybreaks

\usepackage{graphicx}
\newcommand{\be}{\begin{equation}}
\newcommand{\ee}{\end{equation}}
\newcommand{\bea}{\begin{eqnarray}}
\newcommand{\eea}{\end{eqnarray}}
\newcommand{\mc}{\mathcal}
\newcommand{\olambda}{\overline{\lambda}}

\usepackage{subcaption}
\usepackage[colorlinks=true]{hyperref}
\definecolor{wrongultramarine}{rgb}{1,0.5,0}

\definecolor{maria}{RGB}{84,39,143}

\begin{document}

\title{Equilibrium dynamics of infinite-range quantum spin glasses in a field}

\begin{abstract} 
We determine the low-energy spectrum and Parisi replica symmetry breaking function for the spin glass phase of the quantum Ising model with infinite-range random exchange interactions and transverse and longitudinal ($h$) fields. We show that, for all $h$, the spin glass state has full replica symmetry breaking, and 
the local spin spectrum is gapless with a spectral density which vanishes linearly with frequency.   
These results are obtained using an action functional---argued to yield exact results at low frequencies---that expands in powers of a spin glass order parameter, which is is bilocal in time, and a matrix in replica space. We also present the exact solution of the infinite-range spherical quantum $p$-rotor model at nonzero $h$: here, the spin glass state has  one-step replica symmetry breaking, and gaplessness only appears after imposition of an additional marginal stability condition.
Possible connections to experiments on random arrays of trapped Rydberg atoms are noted.
\end{abstract}

\author{Maria Tikhanovskaya}
\affiliation{Department of Physics, Harvard University, Cambridge MA 02138, USA}

\author{Subir Sachdev}
\affiliation{Department of Physics, Harvard University, Cambridge MA 02138, USA}

\author{Rhine Samajdar}
\affiliation{Department of Physics, Princeton University, Princeton, NJ 08544, USA}
\affiliation{Princeton Center for Theoretical Science, Princeton University, Princeton, NJ 08544, USA}

\maketitle
\newpage
\linespread{1.05}
\tableofcontents

\section{Introduction}
Modern advances in the development and control of programmable quantum simulators \cite{MIS22, Dwave-sg, IsingQC1, IsingQC2} have led to remarkable implementations of the idea of solving classical optimization problems by quantum tunnelling \cite{Aeppli99,Farhi01}. In a recent experiment, for instance, \citet{MIS22} used a two-dimensional Rydberg atom array to investigate quantum optimization algorithms and demonstrated a superlinear quantum speedup in finding exact solutions.

In such a setup, each atom can be either in the atomic ground state or in a highly excited Rydberg state (with a large principal quantum number), thus realizing a quantum two-level system, which can be represented by the eigenstates of the Pauli operator $Z_i$ on atom $i$. The laser-induced Rabi flipping is then described by the operator $g \sum_i X_i$, while the laser detuning is given by $\Delta \sum_i Z_i$. The long-ranged van der Waals interactions between two atoms is active only when both are in the Rydberg state, so an interaction $J_{ij}$ between atoms $i$ and $j$, $J_{ij} (1 + Z_i)(1+Z_j)/4$, provides a route to implementing pairwise constraints on the optimization problems \cite{MIS22,kim2022rydberg,ahn_PRXQ,Pichler22,jeong2023quantum}. The atoms, which are trapped in optical tweezers, can be arranged in arbitrary geometries, and positioning them on randomly site-diluted lattices---as in Ref.~\onlinecite{MIS22}---introduces an element of spatial disorder in the Hamiltonian.

Inspired by the success of random, infinite-range models in understanding classical optimization problems \cite{Mezard09}, we will examine here the equilibrium dynamics of the infinite-range Ising spin glass in a field, $h$, with the Hamiltonian 
\begin{align}
    H = \sum_{i<j} J_{ij} Z_i Z_j - g \sum_i X_i - h \sum_i Z_i \label{H_Ising}
\end{align}
where $i,j=1 \ldots N$ denote the lattice sites, and the interactions between them, $J_{ij}$, are taken as independent random numbers drawn from the probability distribution
\begin{equation}\label{pj2}
	P(J_{ij})\propto \exp\left[- \frac{N}{2}\frac{J_{ij}^2}{J^2}\right].
\end{equation}
This model has been much studied in the quantum spin glass literature \cite{Yamamoto_1987,Kopec89,Ray89,Usadel90,Huse93,YSR93,RSY95,Grempel98, Kennett01,Rozenberg01,Muller12,Mukherjee15,Mukherjee18,Young17,Kiss23}, but only a few results have been obtained for the model with a  nonzero field, $h \neq 0$ \cite{RSY95,Young17,Kiss23}, which is an essential ingredient in the optimization toolbox of Rydberg quantum simulators. Here, we shall provide exact results for the equilibrium long-time dynamics of the $N=\infty$ model with $h$ nonzero. Our results here are a prelude to the study of the experimental case where the couplings are time-dependent, but nonequilibrium results will not be presented here.

The equilibrium solution for the $N=\infty$ quantum Ising model with independent random interactions requires the self-consistent solution of a (0+1)-dimensional replicated Ising model with long-range interactions \cite{Young17,Kiss23}. Such a model is not exactly solvable, so a full solution at all times requires Monte Carlo simulations. However, it was argued in Ref.~\onlinecite{RSY95} that a Landau-theory-like strategy of expanding the quantum action functional in powers of an appropriately subtracted $Z$ autocorrelation function (which is also a matrix in replica space) can provide the exact form of the long-time correlations in the $N =\infty$ model. Here, we shall implement this strategy for the $h\neq 0$ model, and obtain the low-frequency dynamic spin spectrum, along with the Parisi spin glass order parameter with full replica symmetry breaking.

Given the difficulty in obtaining the exact solution of the $N=\infty$ quantum Ising model at all times, we will also study a cousin of the quantum Ising model which has been the focus of some attention in the literature, but only at zero field: this is the `spherical quantum $p$-rotor model' \cite{Cugliandolo_Grempel,Cugliandolo_Lozano,Kennett01,Anous:2021eqj,Schiro20,Winer:2022ciz}. In the literature, this model has been referred to as a `$p$-spin' model rather than a `$p$-rotor' model. While the distinction between spins and rotors is not important for classical systems, it is crucial for quantum systems. `Spin' usually refers to a quantum degree of freedom whose components do not commute with each other, while the components of a rotor all commute. Accordingly, we will use the `$p$-rotor' terminology in this paper.

As an aside, we also note studies of the Heisenberg spin glass model \cite{Chowdhury:2021qpy,SY92,PG98,GPS00,GPS01,Biroli02,MJR02,MJR03,Henry21,Dumi22,Christos:2021wno}, which will not be considered in the present paper. In the Heisenberg model, the states on each site $i$ have a twofold degeneracy, unlike both the Ising and rotor models considered
here. An important consequence is that there is no trivial paramagnetic state in the Heisenberg model. In contrast, the Ising and rotor models have a gapped paramagnet with a nondegenerate ground state at large $g$.

We define the spherical quantum $p$-rotor model here by an imaginary time ($\tau$) path integral over continuous real rotor/spin coordinates $\sigma_i$, which are the analogs of the discrete $Z_i= \pm 1$ Ising spins. The partition function of this model at an inverse temperature $\beta = 1/T$, and in a field $h$ is
\begin{equation}
\label{eq:PSM}
	Z[J_{i_1 \ldots i_p}]=\int \mathcal{D}\sigma^{}_i (\tau) \;\exp\left[-\int_0^\beta d\tau\, \left(\frac{1}{2g}\dot{\sigma^{}_i}(\tau)\dot{\sigma^{}_i}(\tau)+\sum_{i_1 <  \ldots < i_p } J_{i_1\ldots i_p}\sigma^{}_{i_1}(\tau)\ldots\sigma^{}_{i_p}(\tau) - h \sum_i \sigma^{}_i (\tau)\right)\right],
\end{equation}
with repeated indices summed over, and $\dot{\sigma^{}_i}(\tau) \equiv d \sigma_i /d \tau$. Importantly, in order to keep the energy finite, we equip the rotors with a spherical constraint
\begin{equation}
\label{eq:Constraint}
	\sum_{i=1}^{N}\sigma_i(\tau)\,\sigma_i(\tau)=N \,,
\end{equation}
{\it i.e.\/}, the rotor degrees of freedom lie on an $N$-dimensional sphere of radius $\sqrt{N}$. These rotors interact with $p$-rotor couplings $J_{i_1\ldots i_p}$, and all of these couplings are taken to be independent random variables with distribution 
\begin{equation}\label{pj}
	P(J_{i_1\ldots i_p})\propto \exp\left[- \frac{N^{p-1}}{p!}\frac{J_{i_1\ldots i_p}^2}{J^2}\right],
\end{equation}
where the width of the distribution over couplings is set by the scale $J$. The factor of $N^{p-1}$ ensures that the Hamiltonian is of order $N$, and accordingly, enforces an extensive scaling of the energy and free energy.

Owing to the global nature of the constraint in (\ref{eq:Constraint}), the solution of the spherical $p$-rotor model is simpler than that of the Ising model with a local constraint $Z_i = \pm 1$ on every site $i$. On the other hand, the global constraint also makes the $p$-rotor model a less physical generalization of the Rydberg experiments. The solution of the $p$-rotor model requires analysis of a closed set of self-consistent Schwinger-Dyson equations, in which the self energies are written as polynomials of the Green's functions. Such Schwinger-Dyson equations have been numerically studied earlier at $h=0$ and in imaginary time and frequency \cite{Cugliandolo_Grempel,Anous:2021eqj}; here, we will extend the numerical solution to $h \neq 0$ and obtain the full dynamic rotor spectrum by direct solution in real-frequency space.
We will also compute the Parisi spin glass order parameter and find that as in the spherical classical $p$-rotor model \cite{Gross84,Gardner85,Crisanti_statics,CC05,Talagrand} (obtained by taking the $g=0$ limit of (\ref{eq:PSM})), there is only one-step replica symmetry breaking.

\subsection{Main results}
\label{sec:main_results}

Our results are expressed in terms of a field $Q_{ab} (\tau_1, \tau_2)$ which is bilocal in imaginary time and a matrix in replica space with indices $a,b =1,\ldots, n$. This field is related to the spin/rotor autocorrelation functions via
\begin{align}
    Q_{ab} (\tau_1, \tau_2) \sim  
    \left\{
    \begin{array}{ccc}
    \displaystyle \frac{1}{N} \sum_{i} Z_{ia} (\tau_1) Z_{ib} (\tau_2) &,& \mbox{Ising model}\\
   \displaystyle \frac{1}{N} \sum_{i} \sigma_{ia} (\tau_1) \sigma_{ib} (\tau_2) &,& \mbox{$p$-rotor model}
    \end{array},
    \right.
\end{align}
where $a,b$ are replica indices.
Time-translational symmetry requires that the $N=\infty$ solution take the form
\begin{equation}
Q_{ab} (\tau_1 , \tau_2 ) = \frac{1}{\beta} \sum_{\nu_n} Q_{ab}
(i \nu_n) e^{i \nu_n (\tau_1 - \tau_2 )}\,, \label{rsy3}
\end{equation}
where $\nu_n$ is a bosonic Matsubara frequency.
We choose the following ansatz for
$Q_{ab}$:
\begin{equation}
Q_{ab} (i \nu_n) = \left\{ \begin{array}{ccc} Q_r (i \nu_n) + \beta
q_{EA} \delta_{\nu_n,0},
& \qquad & a=b \\
\beta q_{ab}  \delta_{\nu_n,0}, & \qquad & a \neq b
\end{array}.
\right. \label{rsy4}
\end{equation}
The replica off-diagonal terms in $Q_{ab}$ are chosen to be time-independent, because there is no correlation between the time evolution of the spins in distinct replicas \cite{RSY95}. We have parameterized these off-diagonal terms in terms of a Parisi matrix $q_{ab}$; as in the classical spin glass theory \cite{HertzBook}, this matrix has an ultrametric structure which is 
characterized by the Parisi function $q(u)$, with $q(1) \equiv
q_{EA}$, the Edwards-Anderson order parameter at $T=0$. We have included an additive factor of $\beta q_{EA}$ in
the replica diagonal term of (\ref{rsy4}) for convenience. We will find that this ensures the solution for $Q_r (\tau)$ vanishes as $\tau \rightarrow \infty$ at $T=0$. Also, the
diagonal components $q_{aa}$ do not appear in the above, and we use this freedom to choose $q_{aa} = 0$. 

\begin{figure}
\includegraphics[width=3in]{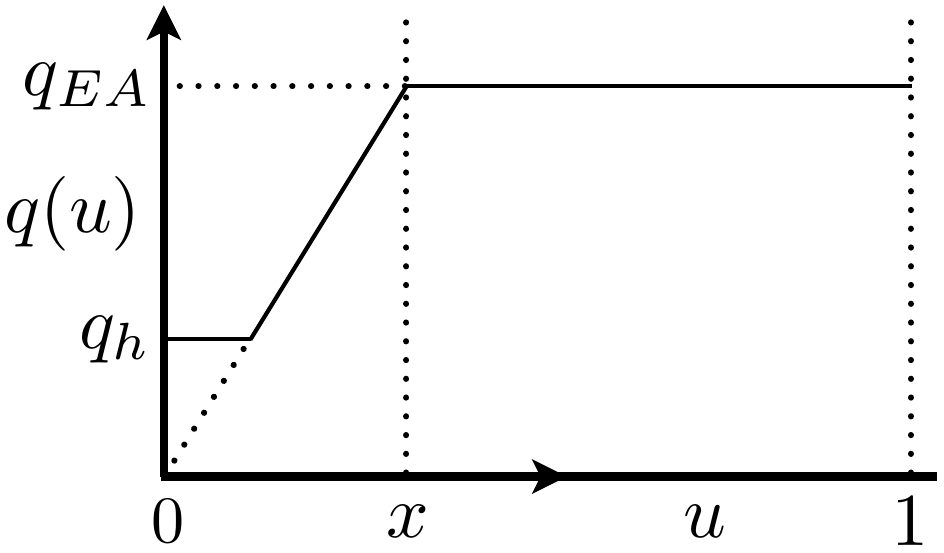}
\caption{The Parisi spin glass order parameter for the Ising model at $h \neq 0$.
}
\label{fig:qu2}
\end{figure}

For the Ising model studied in Sec.~\ref{sec:isingmodel}, we find that low-frequency ($\omega$) spectrum in the replica-symmetry-breaking phase with $h \neq 0$ is the same as that obtained earlier \cite{RSY95,Muller12,Kiss23} at $h=0$:
\begin{align}
\mbox{Im}\, Q_r (\omega) \sim \omega .
\label{Qrlow}
\end{align}
Note that a gapless spectrum for $h \neq 0$ is more surprising than at $h=0$ because there is no $Z_i \rightarrow -Z_i$ symmetry that can be broken in the spin glass phase. The replica symmetry breaking is characterized by the Parisi function $q(u)$ shown in Fig.~\ref{fig:qu2}.
This function has the same form as that found for the classical Ising model with $g=0$ \cite{Thouless80}. The function $q(u)$ is nonanalytic at $u=x$, and the value of $x$ vanishes linearly with $T$ as $T \rightarrow 0$ as (see Eq.~(\ref{rsy8}))
\begin{align}
    x = \frac{2y}{\kappa}\, T\, q_{EA},
\end{align}
where $y$ and $\kappa$ are couplings of order unity  in the Landau action (see Eq.~(\ref{rsy2})).
Thus, although the replica symmetry breaking vanishes at $T=0$, it is nevertheless important that the $T$- and $h$-dependent structure in Fig.~\ref{fig:qu2} be included to obtain the gapless spectrum at $T=0$ in (\ref{Qrlow}).
The smaller $u$ plateau in Fig.~\ref{fig:qu2} is at \cite{Thouless80,RSY95}
\begin{align}
q_h = \left( \frac{3h^2}{8 y} \right)^{1/3} \,. \label{quh1}
\end{align}

\begin{figure}
\includegraphics[width=3in]{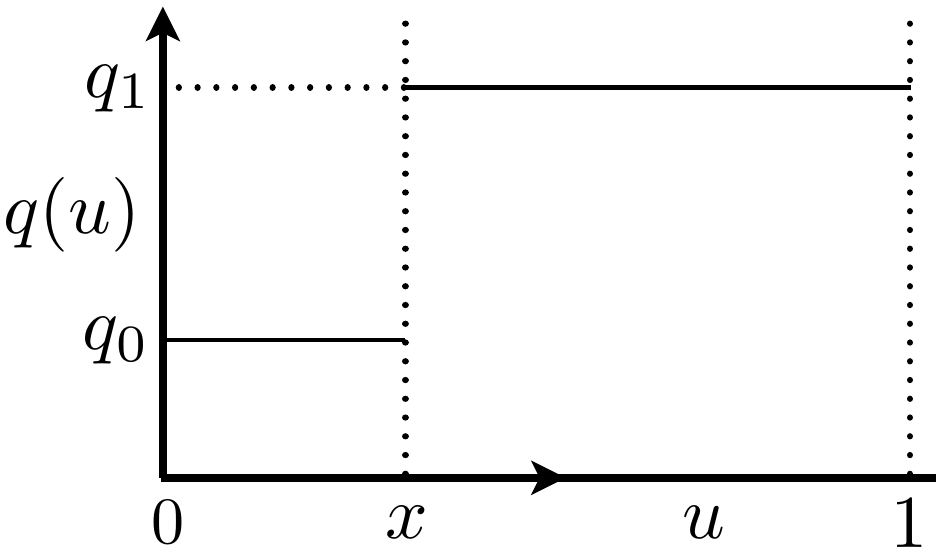}
\caption{The Parisi spin glass order parameter for the spherical quantum $p$-rotor model for $p \geq 3$.
}
\label{fig:qup}
\end{figure}

\begin{figure}
\center{
\includegraphics[width=0.5\linewidth]{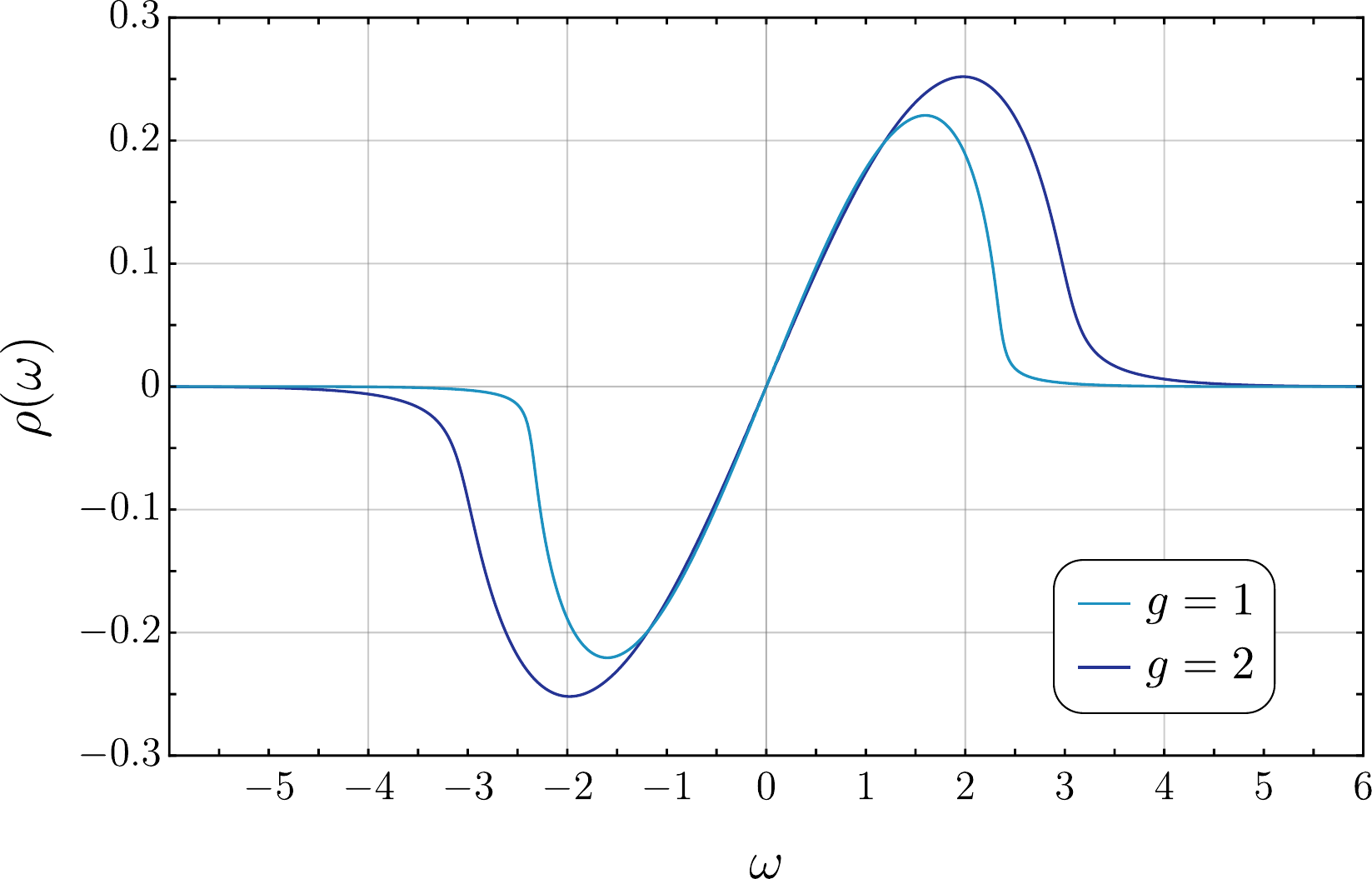}}
\caption{Behavior of the zero-temperature spectral function $\rho (\omega) = \mbox{Im} Q_r (\omega)/\pi$, as determined by Eqs.~\eqref{eq:sp_dens_RSB1}--\eqref{eq:sp_dens_RSB2}, for the one-step replica-symmetry-breaking solution of the $p=3$ spherical quantum $p$-rotor model at different values of the transverse field $g$ (with $J=1$). The marginal stability condition (\ref{gps4}) is imposed. The spectrum is independent of $h$ as long as we are in the replica-symmetry-breaking phase in Fig.~\ref{fig:phase_diag_pspin}.}
\label{fig:sp_dens_RSB}
\end{figure}

The results for the spherical quantum $p$-rotor model, $p \geq 3$, studied in Sec.~\ref{sec:spherical_model}, do have differences from the Ising model tied to the one-step replica symmetry breaking in the former, shown in Fig.~\ref{fig:qup}, which is in turn tied to the nonlocal rotor constraint in (\ref{eq:Constraint}).
This $q(u)$ is characterized by a break-point at $u=x$, and the large-$N$ saddle-point equations leave the value of $x$ undetermined. The mathematical solution of the classical problem at $g=0$ \cite{Talagrand} implies we should compute the free energy $\mathcal{F}$, and then use the solution of $\partial \mathcal{F}/\partial x = 0$ to determine $x$. At such an $x$, and indeed at all generic values of $x$, the spectrum of $Q_r (\omega)$ turns out to have a gap. Earlier work \cite{Cugliandolo_Kurchan,Cugliandolo_Lozano,Cugliandolo_Grempel} has advocated use of a `marginal stability' condition, which leads to precisely the value of $x$ for which the spectrum is gapless. We find that such a gapless spectrum obeys (\ref{Qrlow}), and show a plot of $\mbox{Im}\, Q_r (\omega)$
for all frequencies in Fig.~\ref{fig:sp_dens_RSB}.
The lower plateau at $q_0$ in Fig.~\ref{fig:qup} vanishes as $h$\,$\rightarrow$\,$0$ as $q_0$\,$\sim$\,$h^2$, unlike (\ref{quh1}) for the Ising model. The value of $x$ still vanishes linearly with $T$, as in the Ising model.
We note that
the $p$\,$=$\,$2$ spherical model has no replica symmetry breaking, and the rotor spectrum is always gapped for $h \neq 0$.

\section{Ising model}
\label{sec:isingmodel}

\subsection{Landau action}

We motivate the structure of the Landau action for the Ising model by 
recalling the solution of the spherical model for the $p=2$ case \cite{YSR93}, in which the interactions have the same form as in the Ising model, but the local Ising constraint has been replaced by the global constraint in (\ref{eq:Constraint}). At $h=0$, the large-$N$ solution in the paramagnetic phase is of the form (see Appendix~\ref{sec:p2spherical}, and Chapter 33 in Ref.~\onlinecite{QPM})
\begin{align}
Q_{ab} (i \nu_n) =     \frac{2g \delta_{ab}}{\nu_n^2 + (\Delta^2 + \Lambda^2)/2 + \left[ (\nu_n^2 + \Delta^2)(\nu_n^2 + \Lambda^2 )\right]^{1/2}},
\end{align}
where $\nu_n$ is a Matsubara frequency, $\Delta$ is a small energy gap, and $\Lambda$ is a high cutoff frequency. In the spin glass phase of the spherical model, the solution is gapless and replica symmetric:
\begin{align}
    Q_{ab} (i \nu_n) = \beta q_{EA} \delta_{\nu_n,0} +
    \frac{2g \delta_{ab}}{\nu_n^2 + \Lambda^2/2 + |\nu^{}_n| \left[\nu_n^2 + \Lambda^2\right]^{1/2}},
\end{align}
where $q_{EA}$ is the Edwards-Anderson order parameter.

We shall be interested in extending this solution beyond the spherical limit to the Ising model---at low frequency scales in the vicinity of the quantum critical point, where $|\nu_n|$, $\Delta$, $T$ $\ll \Lambda$, $q_{EA}$ is small---and allow for replica symmetry breaking. In this regime, we can approximate the paramagnetic solution of the spherical model by
\begin{align}
Q_{ab} (i \nu_n) =   \delta_{ab} \left[ 
A - B \sqrt{\nu_n^2 + \Delta^2} + \ldots \right]\,, \label{Qform1}
\end{align}
where $A,B$ are positive constants, 
while the spin glass solution is 
\begin{align}
    Q_{ab} (i \nu_n) = \beta q_{EA} \delta_{\nu_n,0} + \delta_{ab} \left[ 
A - B |\nu_n| \right] + \ldots \quad.\label{Qform2}
\end{align}
We now notice that if we shift $Q$ by a frequency-independent and replica-diagonal constant,
\begin{align}
    Q_{ab} (i \nu_n) \rightarrow Q_{ab} (i \nu_n) - A \delta_{ab},
\end{align}
then the shifted $Q_{ab} (i \nu_n)$ becomes small at the relevant low-frequency scales on both sides of the quantum critical point. This makes the shifted $Q_{ab} (i \nu_n)$ a suitable field in which to carry out the Landau expansion of the Ising model. In terms of the original bilocal field, the shift is 
\begin{align}
    Q_{ab} (\tau_1, \tau_2) \rightarrow Q_{ab} (\tau_1, \tau_2) - A \, \delta_{ab} \, \delta(\tau_1 - \tau_2), \label{Qtaushift}
\end{align}
and the constant $A$ characterizes nonuniversal, short-time physics not of interest to us.

Working with this shifted field, the important low-order terms in the Landau expansion for the action of the Ising model are \cite{RSY95}
\begin{align}
\mathcal{A}_h = \mathcal{A} - \frac{h^2}{2} \sum_{ab} \int d \tau_1 d \tau_2 Q_{ab} (\tau_1, \tau_2)  - \frac{\beta}{2} \chi_{hb} h^2, \label{rsy2h}
\end{align}
where $h$ is the longitudinal field, $\chi_{hb}$ is a background contribution to the linear spin susceptibility, and the $h=0$ Landau action is
\begin{eqnarray}
\nonumber
{\mathcal A} &=& \frac{1}{\kappa} \int d \tau \sum_a \left. \left[
\frac{\partial}{\partial \tau_1} \frac{\partial}{\partial \tau_2}
+ r \right] Q_{aa} (\tau_1, \tau_2 ) \right|_{\tau_1 = \tau_2 =
\tau} -\frac{\kappa}{3} \int d \tau_1 d \tau_2 d \tau_3 \sum_{abc}
Q_{ab} (\tau_1 , \tau_2 ) Q_{bc} (\tau_2, \tau_3) Q_{ca} (\tau_3,
\tau_1 ) \nonumber \\
&+&\frac{U}{2} \int d \tau \sum_a Q_{aa} (\tau,\tau)
Q_{aa} (\tau,\tau) - \frac{y}{6} \int d \tau_1 d \tau_2 \sum_{ab}
\left[ Q_{ab} (\tau_1 , \tau_2 ) \right]^4; \label{rsy2}
\end{eqnarray}
see Appendix~A of Ref.~\onlinecite{RSY95} for a derivation of Eq.~\eqref{rsy2} from (\ref{H_Ising}). Here, $r$ is the parameter which tunes across the spin glass
transition at $h=0$: it is analogous to the coupling $g$ in the Ising Hamiltonian in (\ref{H_Ising}). The cubic term $\kappa$ is analogous to the cubic term in Parisi's original theory of classical spin glasses \cite{HertzBook,Parisi79}. The $U$ term only involves a single replica, and is a quantum self-interaction of the soft Ising order parameter. The $y$ term is analogous to a quartic term in the classical case \cite{HertzBook}, where it is the term responsible for full replica symmetry breaking in the spin glass phase; however, the dynamic quantum effects of $y$ are weak, and can be treated perturbatively.

Note that (\ref{rsy2}) does not contain the allowed quadratic term $ \int d \tau_1 d \tau_2 [Q_{ab} (\tau_1, \tau_2)]^2 $. We have removed such a term by exploiting the shift in (\ref{Qtaushift}). We will see below that such a choice is equivalent to the requirement for the validity of the Landau theory that the full function $Q_{ab} ( i \nu_n)$ is small near the quantum critical point. This shift strategy is analogous to that followed for the critical theory of the Yang-Lee edge singularity \cite{MEF_edge,Cardy_edge}.

Our analysis of the physics of the action $\mathcal{A}_h$ will examine the behavior as a function of the tuning parameter across the spin glass transition $r$, the longitudinal field $h$, and the temperature $T$. We will keep the couplings $\kappa$ and $U$ of order unity and obtain results for small $y$. We find that for $h >0$ there are contributions nonanalytic in $y$ that are important to include; however, analytic corrections in integer powers of $y$ are not crucial, and these are relegated to Appendix~\ref{app:y1}.

The analysis of thermodynamic properties in the
spin glass phase  in Ref.~\onlinecite{RSY95} was carried out with a
vanishing coefficient of the quartic term, $y=0$: in this case
the order parameter has replica symmetry. Here, we will
extend the solution to small $y \neq 0$, and show that the
solution has broken replica symmetry. We also show that the spin fluctuation spectrum remains gapless both at $y=0$ and $y \neq 0$.

\subsection{Free energy}

Inserting the time-translational symmetric ansatz of (\ref{rsy3}) into (\ref{rsy2h}) and (\ref{rsy2}), we obtain the free energy 
\begin{align}
\mathcal{F}_h = \mathcal{F} - \frac{h^2}{2} \sum_{ab} Q_{ab} (i\nu_n = 0)  - \frac{1}{2} \chi_{hb} h^2\,,
\end{align}
where the $h$-independent free energy is
\begin{align}
\mathcal{F} &= \frac{1}{\beta \kappa} \sum_a \sum_{\nu_n}  ( \nu_n^2 + r) Q_{aa} (i
\nu^{}_n) - \frac{\kappa}{3 \beta} \sum_{abc} \sum_{\nu_n}  Q_{ab} (i \nu^{}_n) Q_{bc} (i \nu^{}_n) Q_{ca} (i \nu^{}_n) +
\frac{U}{2} \sum_a \left[ \frac{1}{\beta} \sum_{\nu_n} Q_{aa} ( i\nu^{}_n ) \right]^2 \nonumber \\
&- \frac{ y}{6 \beta^3} \sum_{ab} \sum_{\nu_n, \nu^{\prime}_n,
\nu^{\prime\prime}_n} Q_{ab} (i\nu^{}_n) Q_{ab} (i\nu^{\prime}_n)
Q_{ab} (i\nu^{\prime\prime}_n) Q_{ab} (- i\nu^{}_n -
i\nu^{\prime}_n-i\nu^{\prime\prime}_n). \label{rsy3a}
\end{align}
We now insert (\ref{rsy4}) into (\ref{rsy3a}) and obtain the free energy
\begin{align}
\frac{\mathcal{F}_h}{n} = \mathcal{F}_{sg,h} + \mathcal{F}_{Q} \label{FFF},
\end{align}
which consists of two pieces: a `spin glass' and a `quantum' component. 

The first `spin glass' component in (\ref{FFF}) is given by
\begin{align}
\mathcal{F}_{sg,h} = \mathcal{F}_{sg} - \frac{\beta h^2}{2n} \sum_{ab} q_{ab} - \frac{h^2}{2} Q_r (0) - \frac{\beta h^2}{2} q_{EA} - \frac{1}{2} \chi_{hb} h^2\,,\label{rsy5h}
\end{align}
where the $h$-independent terms are
\begin{align}
\mathcal{F}_{sg} = &  - R_1 \frac{1}{n} \mbox{Tr} q^2 - \frac{R_2}{3} \frac{1}{n} \mbox{Tr} q^3
- \frac{R_3}{6} \frac{1}{n} \sum_{ab} q_{ab}^4 \nonumber \\
& + r \frac{q_{EA}}{\kappa} - \frac{\kappa}{3 \beta} \left[ (Q_r (0) + \beta q_{EA})^3 - Q_r(0)^3 \right] - \frac{\beta y}{6} q_{EA}^4 - \frac{2y}{3} q_{EA}^3 Q_r (0)\,, \label{rsy5}
\end{align}
with
\begin{eqnarray}
R_1 &=& \beta \kappa (Q_r (0) + \beta q_{EA}),\quad
R_2 = \kappa \beta^2, \quad
R_3 = \beta y. \label{rsy6}
\end{eqnarray}
The terms involving $q_{ab}$ in (\ref{rsy5}) are identical to those the Landau theory of the classical spin glass in Section 3.4 of Ref.~\onlinecite{HertzBook}.

The second `quantum' component in (\ref{FFF}) is $h$-independent
\begin{eqnarray}
\mathcal{F}_Q &=&   \frac{1}{\beta \kappa} \sum_{\nu_n} ( \nu_n^2 + r) Q_r(i
\nu^{}_n) - \frac{\kappa}{3 \beta} \sum_{\nu_n} Q_r^3 (i \nu^{}_n) +
\frac{U}{2} \left[ \frac{1}{\beta} \sum_{\nu_n} Q_r ( i\nu^{}_n ) +
q_{EA} \right]^2 \nonumber \\
&-&  \frac{y q_{EA}^2}{\beta} \sum_{\nu_n} Q_r(i\nu^{}_n) Q_r(-i\nu^{}_n)
- \frac{2 y q_{EA}}{3 \beta^2} \sum_{\nu_n, \nu^{\prime}_n}
Q_r (i\nu^{}_n) Q_r (i\nu^{\prime}_n) Q_r(- i\nu^{}_n - i\nu^{\prime}_n)
\nonumber
\\ &-& \frac{ y}{6 \beta^3} \sum_{\nu_n, \nu^{\prime}_n,
\nu^{\prime\prime}_n} Q_r (i\nu^{}_n) Q_r (i\nu^{\prime}_n)
Q_r (i\nu^{\prime\prime}_n) Q_r (- i\nu^{}_n -
i\nu^{\prime}_n-i\nu^{\prime\prime}_n). \label{FQM}
\end{eqnarray}
We will solve the saddle-point equations for $\mathcal{F}_{sg}$ exactly, while those for $\mathcal{F}_Q$ can be solved order-by-order in $y$. We will present the $y^0$ solution below, while the $y^1$ solution is presented in Appendix~\ref{app:y1}. It turns out that the solution of $\mathcal{F}_{sg}$ contains terms nonperturbative in $y$ for $h \neq 0$,  so it is important to treat the spin glass terms exactly.

\subsection{Saddle-point equations}

The saddle-point equations are most easily determined by taking the derivative of (\ref{rsy3a}) with respect to $Q_{ab}( i \nu_n)$:
\begin{align}
\frac{\beta h^2}{2} \delta_{\nu_n,0} &= \frac{1}{\kappa}  ( \nu_n^2 + r)  \delta_{ab}  - \kappa \sum_{c}   Q_{ac} (i \nu^{}_n) Q_{cb} (i \nu^{}_n) +
U \delta_{ab} \frac{1}{\beta} \sum_{\nu_n^\prime} Q_{aa} ( i\nu_n^\prime )  \nonumber \\
&- \frac{ 2y}{3 \beta^2}  \sum_{ \nu^{\prime}_n,
\nu^{\prime\prime}_n} Q_{ab} (i\nu^{\prime}_n)
Q_{ab} (i\nu^{\prime\prime}_n) Q_{ab} (- i\nu^{}_n -
i\nu^{\prime}_n-i\nu^{\prime\prime}_n). \label{rsy3bb}
\end{align}
The replica off-diagonal equation of (\ref{rsy3bb}) is
\begin{align}
\frac{\beta h^2}{2} = - 2 R_1 q_{ab} - R_2 \sum_c q_{ac} q_{cb} - \frac{2 R_3}{3} q_{ab}^3\,, \label{rsy3cc}
\end{align}
while the replica diagonal part gives
\begin{align}
0 &= \frac{1}{\kappa}  ( \nu_n^2 + r)  - \kappa \left[Q_r (i \nu^{}_n) \right]^2 +
 \frac{U}{\beta} \sum_{\nu_n^\prime} Q_r ( i\nu_n^\prime ) + u q_{EA}  \nonumber \\
&- \frac{ 2y}{3 \beta^2}  \sum_{ \nu^{\prime}_n,
\nu^{\prime\prime}_n} Q_{r} (i\nu^{\prime}_n)
Q_{r} (i\nu^{\prime\prime}_n) Q_{r} (- i\nu^{}_n -
i\nu^{\prime}_n-i\nu^{\prime\prime}_n) \nonumber \\
&- \frac{2 y}{\beta} q_{EA} \sum_{\nu_n^\prime} Q_r (i\nu_n^\prime) Q_r( i\nu^{}_n - i\nu_n^\prime) - 
2 y q_{EA}^2 Q_r (i \nu^{}_n) \nonumber \\
&+  \delta_{\nu_n,0}\left[ - \kappa\beta^2 \sum_c q^{}_{ac}q^{}_{ca} - \kappa \beta^2 q_{EA}^2 -2 \kappa\beta q^{}_{EA} Q_r (0) - \frac{2 \beta y}{3} q_{EA}^3 - \frac{\beta h^2}{2} \right] \,.
\label{rsy3dd}
\end{align}

\subsection{Zero-field limit}
\label{sec:0}

In this section, we first rederive the results obtained earlier  \cite{RSY95,GPS01} at $h$\,$=$\,$0$, which we will then contrast with the case for a nonzero field later in Sec.~\ref{sec:h}.

\subsubsection{Quantum paramagnet}
\label{sec:0para}

In the paramagnetic phase, with $q_{EA} = 0$ and $q_{ab}=0$, 
the spin glass free energy is $\mathcal{F}_{sg} = 0$. At $h=0$, $\mathcal{F}_{sg,h}=0$ wherefore only the quantum component survives, and Eq.~(\ref{rsy3dd}) simplifies to an equation for $Q_r (i \nu_n)$ alone:
\begin{align}
0 &= \frac{1}{\kappa}  ( \nu_n^2 + r)  - \kappa \left[Q_r (i \nu^{}_n) \right]^2 +
 \frac{U}{\beta} \sum_{\nu_n^\prime} Q_r ( i\nu_n^\prime )   - \frac{ 2y}{3 \beta^2}  \sum_{ \nu^{\prime}_n,
\nu^{\prime\prime}_n} Q_{r} (i\nu^{\prime}_n)
Q_{r} (i\nu^{\prime\prime}_n) Q_{r} (- i\nu^{}_n -
i\nu^{\prime}_n-i\nu^{\prime\prime}_n)  \,.
\label{rsy3d1}
\end{align}
This can be solved iteratively in powers of $y$. At order $y^0$, the solution agrees with the form in (\ref{Qform1})
\begin{align}
Q_{r0} (i \nu_n) = - \frac{\sqrt{\nu_n^2 + \Delta^2}}{\kappa}\,, \label{p1}
\end{align}
where the gap $\Delta$ is given by the solution of 
\begin{align}
\Delta^2 = r - \frac{U}{\beta} \sum_{|\nu_n| < \Lambda} \sqrt{\nu_n^2 + \Delta^2}, \label{p2}
\end{align}
and we have introduced a frequency cutoff to make the frequency summation finite. We will also need the cutoff at higher orders in $y$, but the low-frequency dynamics should remain cutoff-independent.
The critical point $r=r_{c0}$ is obtained by setting $\Delta=0$, and is given by
\begin{align}
r_{c0} = \frac{U}{\beta} \sum_{|\nu_n| < \Lambda} |\nu_n|\,. \label{p3}
\end{align}
The order $y^1$ corrections to the saddle point appear in Appendix~\ref{app:0para}.

\paragraph{Free energy:}

As mentioned above, for the case of the paramagnet, we have $\mathcal{F}_{sg} =0$ for the spin glass component of the free energy  in (\ref{FFF}).
The free energy $\mathcal{F}_Q$ in (\ref{FFF}) at order $y^0$ is 
\begin{align}
    \mathcal{F}_Q^0 = - \frac{2}{3 \kappa^2 \beta} \sum_{\nu_n} (\nu_n^2 + \Delta^2)^{3/2} - \frac{\left[ \Delta^2 -r \right]^2}{2 \kappa^2 u}  \,, \label{e100}
\end{align}
where $\Delta$ is the solution of (\ref{p2}).
The order $y^1$ correction to $\mathcal{F}_Q$ appears in Appendix~\ref{app:0para}.

\subsubsection{Spin glass}
\label{sec:0sg}

We begin by solving (\ref{rsy3cc}) at $h$\,$=$\,$0$, which is the same equation of state as that obtained for the classical spin glass. In terms of the Parisi function $q(u)$ characterizing the $n$\,$\rightarrow$\,$0$ limit of the replica matrix $q_{ab}$, we can write (\ref{rsy3cc}) at $h=0$ as (see Appendix~\ref{app:replica})
\begin{align}
2 R_1 q(u) + \frac{2 R_3}{3} q^3 (u) = R_2 \left[ 2 q(u) \int_0^1 q (v) dv + \int_0^u dv (q(u) - q(v))^2 \right]. \label{par1}
\end{align}
\begin{figure}[tb]
\includegraphics[width=3in]{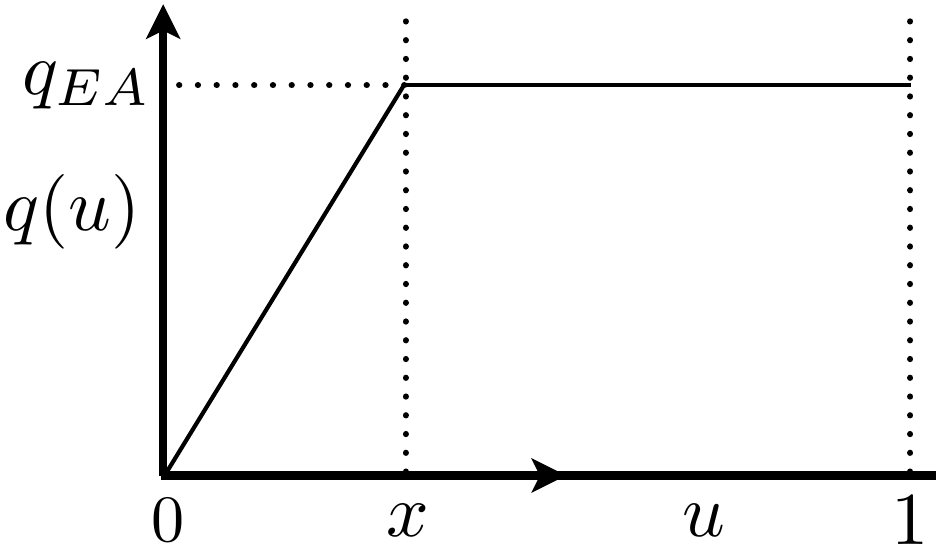}
\caption{The Parisi spin glass order parameter for the Ising model at $h=0$.
}
\label{fig:qu1}
\end{figure}
This is the same as Eq.~(3.74) in Ref.~\onlinecite{HertzBook} for the classical case. A replica-symmetric solution with $q(u) =$ constant is possible, but this is not the preferred solution at $h=0$ (as in the classical case, and, as we will see in Sec.~\ref{sec:h}, for the quantum $h\ne 0$ case). So, we only consider the replica-symmetry-breaking solution here, as shown in Fig.~\ref{fig:qu1}:
\begin{align}
q(u) = \left\{ \begin{array}{ccc} 
q_{EA} u/x & , & 0 < u < x \\
q_{EA} &, & x < u < 1
\end{array}, \right.
\label{qu1}
\end{align}
with
\begin{align}
q_{EA} & = \frac{R^{}_2 - (R_2^2 - 4 R^{}_1 R^{}_3)^{1/2}}{2 R_3}, \nonumber \\
x & = \frac{2 R_3 q_{EA}}{R_2} \,. \label{hf1}
\end{align}
Using (\ref{rsy6}), we can exactly simplify (\ref{hf1}) to
\begin{equation}
Q_r (0) = - \frac{y}{\kappa} q_{EA}^2, \quad x = 2 y q_{EA} /(\beta \kappa) \,. \label{rsy8}
\end{equation}
It can now be verified that the term proportional to $\delta_{\nu_n, 0}$ in (\ref{rsy3dd}) vanishes
\begin{align}
  \kappa\beta^2 \int_0^1 du [q(u)]^2 - \kappa \beta^2 q_{EA}^2 -2 \kappa\beta q^{}_{EA} Q_r (0) - \frac{2 \beta y}{3} q_{EA}^3  = 0\,.
\label{rsy3e}
\end{align}
We can therefore conclude that the complete saddle-point equations for $Q_r (i \nu_n)$ and $q(u)$ reduce to the following three equations for $Q_r (i \nu_n)$, $q_{EA}$ and $x$:
\begin{align}
0 &= \frac{1}{\kappa}  ( \nu_n^2 + r)  - \kappa \left[Q_r (i \nu^{}_n) \right]^2 +
 \frac{U}{\beta} \sum_{\nu_n^\prime} Q_r ( i\nu_n^\prime ) + u q_{EA}  \nonumber \\
 &- \frac{ 2y}{3 \beta^2}  \sum_{ \nu^{\prime}_n,
\nu^{\prime\prime}_n} Q_{r} (i\nu^{\prime}_n)
Q_{r} (i\nu^{\prime\prime}_n) Q_{r} (- i\nu^{}_n -
i\nu^{\prime}_n-i\nu^{\prime\prime}_n) \nonumber \\
& - \frac{2 y}{\beta} q^{}_{EA} \sum_{\nu_n^\prime} Q_r (i\nu_n^\prime) Q_r( i\nu^{}_n - i\nu_n^\prime) - 
2 y q_{EA}^2 Q_r (i\nu^{}_n), \label{ifinaleq1} \\
y q_{EA}^2 & = - \kappa Q_r (0), \label{ifinaleq2} \\
\beta \kappa x & = 2 y q_{EA}. \label{ifinaleq3}
\end{align}

\paragraph{Gapless condition:}

Now, let us analytically continue (\ref{ifinaleq1}) to real frequency, and assume that
\begin{align}
Q_r (\omega \rightarrow 0) = Q_r (0) + |\omega|^{\alpha} \left[ a + i b\, \mbox{sgn}(\omega) \right]
\end{align}
with $\alpha > 0$.
Then, at $T=0$, the terms of order $Q_r^2 $ and $Q_r^3$ in (\ref{ifinaleq1}) will have imaginary parts which vanish faster than $|\omega|^\alpha$ as $|\omega| \rightarrow 0$. Collecting all terms of order  $|\omega|^\alpha$ in the imaginary part of (\ref{ifinaleq1}), we obtain the condition
\begin{align}
-2 \kappa Q_r (0) - 2 y q_{EA}^2 = 0, \label{gapless}
\end{align}
which is automatically satisfied from (\ref{ifinaleq2}). So, the spin dynamics are gapless at all values of $y$ in the spin glass phase with $h=0$. We will see in the explicit solution below that the exponent $\alpha=1$, so that $\mbox{Im} Q_r (\omega \rightarrow 0) \sim \omega$. 

\paragraph{Solution:}
\label{sec:solution}

The equations (\ref{ifinaleq1}--\ref{ifinaleq3}) can be solved iteratively in powers of $y$.
At order $y^0$, we have
\begin{align}
q_{EA0} &= \frac{1}{\beta \kappa} \sum_{\nu_n} |\nu_n| -
\frac{r}{\kappa U}  \equiv \frac{1}{\kappa U} (r_{c0} - r), \nonumber \\
Q_{r0} (i\nu_n) &= - \frac{|\nu_n|}{\kappa}, \nonumber \\
x &= 0\,.
\end{align}
The order $y^1$ corrections to the saddle point appear in Appendix~\ref{app:0sg}.

\paragraph{Free energy:}
\label{sec:freersb}

Inserting the solution (\ref{qu1})--(\ref{rsy8}) back into (\ref{rsy5}), and using the relation (see Appendix~\ref{app:replica})
\begin{align}
 &   - R_1 \frac{1}{n} \mbox{Tr} q^2 - \frac{R_2}{3} \frac{1}{n} \mbox{Tr} q^3
- \frac{R_3}{6} \frac{1}{n} \sum_{ab} q_{ab}^4 \nonumber \\
&= R_1 \int_0^1 du [q(u)]^2 - \frac{R_2}{3} \int_0^1 du \left[ u [q(u)]^3 + 3 q(u) \int_0^u dv [q(v)]^2 \right] + \frac{R_3}{6}\int_0^1 du [q(u)]^4,
\end{align}
we obtain
\begin{align}
   \mathcal{F}_{sg} = \frac{r q^{}_{EA}}{\kappa} + \frac{y^2 q_{EA}^5}{5\kappa}\,. \label{Fsg0}
\end{align}
This is the full expression for $\mathcal{F}_{sg}$ in (\ref{FFF}), valid to all orders in $y$ in terms of the exact $q_{EA}$. However, $q_{EA}$ is known only to order $y^1$ in (\ref{qea11}).

As in the paramagnet, we will only determine $\mathcal{F}_Q$ in (\ref{FFF}) order-by-order in $y$. At order $y^0$, the value of $\mathcal{F}_Q$ is
\begin{align}
\mathcal{F}_Q^0 = - \frac{r q_{EA0}}{\kappa} - \frac{2}{3 \kappa^2 \beta} \sum_{\nu_n} |\nu_n|^{3/2} - \frac{r^2}{2 U \kappa^2}. \label{Fq0sg} 
\end{align}
The order $y^1$ correction to $\mathcal{F}_Q$ appears in Appendix~\ref{app:0sg}.

\subsubsection{Phase diagram}

At order $y^0$, the system is the spin glass phase for $r$\,$<$\,$r_{0c}$, where $r_{c0}$ is given by Eq.~(\ref{p3}).
In the classical limit, $T \gg \Lambda$, we need only consider the $\nu_n = 0$ term, so the phase boundary is at $r = 0$.
In the quantum limit $T \ll \Lambda$, we can evaluate the summation over Matsubara frequencies using the identity
\begin{align}
\frac{1}{\beta} \sum_{\nu_n} D(i \nu_n) = \int_0^{\infty} \frac{ d \nu}{\pi} D(i \nu) + 2 \int_0^{\infty} \frac{d \Omega}{\pi} \frac{\mbox{Im} D(\Omega)}{e^{\Omega/T} - 1} \label{ebose}
\end{align}
for odd spectral functions $\mbox{Im}D(\Omega) = - \mbox{Im}D(-\Omega)$.
Then, we obtain the phase boundary at 
\begin{align}
r  = U \frac{\Lambda^2}{2 \pi} - U \frac{\pi T^2}{3}\,. \label{p3a}
\end{align}

\subsection{Nonzero field}
\label{sec:h}

Having reviewed the results in the absence of a longitudinal field above, we now address the $h\ne 0$ case and determine its phase diagram.

\subsubsection{Replica-symmetric solution}
\label{sec:hrs}

Due to the longitudinal magnetic field, there is an average moment on each site, so $q_{EA}$ is always nonzero; therefore, the replica-symmetric solution is nothing but the paramagnet. For $q_{ab} = q_{EA} (1 - \delta_{ab})$, Eq.~(\ref{rsy3cc}), together with (\ref{rsy6}), reads
\begin{align}
3h^2 + 12 \kappa q^{}_{EA} Q_r (0) + 4y q_{EA}^3  = 0 \,.\label{rs1}
\end{align}

Resultantly, the term proportional to $\delta_{\nu_n,0}$ in (\ref{rsy3dd}) vanishes so that the equation for $Q_r (i \nu_n)$ remains unchanged from that in (\ref{ifinaleq1}).
The gapless condition in  (\ref{gapless}) {\it is not\/} satisfied by (\ref{rs1}), even though $q_{EA} \neq 0$. Thus, the solution for $Q_r (i \nu)$
will have a gap.

At order $y^0$, the solution for $Q_{r0} (i \nu_n)$ from (\ref{rsy3dd}) has the same form as (\ref{p1}), but the equation for the gap $\Delta$ in (\ref{p2}) is now modified to \cite{RSY95}
\begin{align}
\Delta^2 = r + \frac{U\kappa h^2}{4 \Delta} - \frac{U}{\beta} \sum_{|\nu_n| < \Lambda} \sqrt{\nu_n^2 + \Delta^2}\, \label{rs2}
\end{align}
while
\begin{align}
    q_{EA0} = \frac{h^2}{4 \Delta}\,. \label{qeaDelta}
\end{align}
The solutions to Eq.~(\ref{rs2}) are shown in Fig.~\ref{fig:LTDelta}.
The order $y^1$ corrections to the saddle point appear in Appendix~\ref{app:hrs}.

\begin{figure*}
    \centering
    \includegraphics[width=\linewidth]{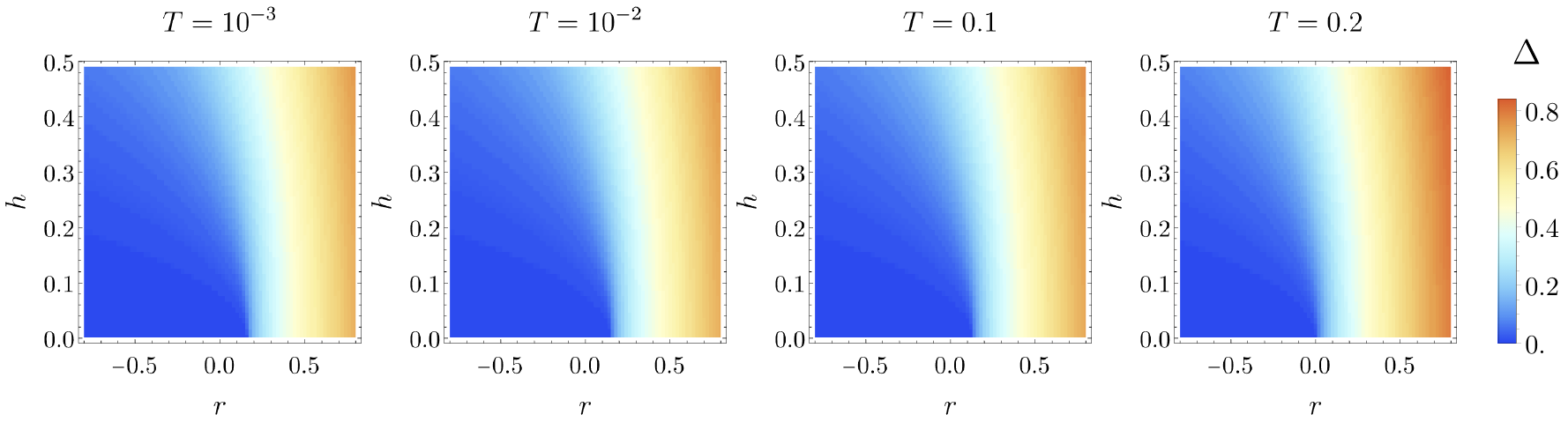}
    \caption{The gap $\Delta$ of the replica-symmetric solution of the Ising model as a function of $r$ and $h$ for various temperatures, as determined from \eqref{rs2}. The parameters $U$, $\kappa$, and $\Lambda$ have all been set to unity in these calculations.}
    \label{fig:LTDelta}
\end{figure*}

\paragraph{Free energy:}

Inserting the solution for $Q_r (0)$ in (\ref{rs1}) into 
(\ref{rsy5h}) we obtain the spin glass component of the free energy
\begin{align}
    \mathcal{F}_{sg,h} = \frac{q^{}_{EA} r}{\kappa} + \frac{y^2 q_{EA}^5}{9 \kappa} + \frac{h^2 q_{EA}^2 y}{6 \kappa} + \frac{h^4}{16 q^{}_{EA} \kappa} - \frac{1}{2} \chi_{hb} h^2. \label{rsy5rs}
\end{align}
Note that we need to insert the solution for $q_{EA}$ in (\ref{Qr1a}) into (\ref{rsy5rs}).

For the quantum component, the order $y^0$ contribution to the free energy in (\ref{FQM}) is
\begin{align}
    \mathcal{F}_{Q}^0 = - \frac{2}{3 \kappa^2 \beta} \sum_{\nu_n} (\nu_n^2 + \Delta^2)^{3/2} - \frac{\left[ \Delta^2 -r \right]^2}{2 \kappa^2 U}  - \frac{h^2 (r-\Delta^2)}{4\kappa \Delta}\,, \label{fq0}
\end{align}
which reduces to (\ref{e100}) at $h=0$.
The order $y^1$ correction to $\mathcal{F}_Q$ appears in Appendix~\ref{app:hrs}.

\subsubsection{Replica symmetry breaking}
\label{sec:rsb:hh}

From (\ref{rsy3cc}), the equation (\ref{par1}) for the Parisi function is modified to
\begin{align}
\frac{\beta h^2}{2} + 2 R_1 q(u) + \frac{2 R_3}{3} q^3 (u) = R_2 \left[ 2 q(u) \int_0^1 q (v) dv + \int_0^u dv (q(u) - q(v))^2 \right]. \label{par2}
\end{align}
The solution of (\ref{par2}) is modified from (\ref{qu1}) to that shown in Fig.~\ref{fig:qu2}
\begin{align}
q(u) = \begin{cases}
q_h, &  0 < u < (q_h/q_{EA}) x \\
q_{EA} u/x, & (q_h/q_{EA}) x < u < x \\
q_{EA}, & x < u < 1
\end{cases} \,. \label{quh}
\end{align}
The function $q(u)$ is nonanalytic at $u=(q_h/q_{EA})x$ and $u=x$, but has no discontinuities. The values of $q_{EA}$ and $x$ are unchanged from those in (\ref{rsy8}), and the value of $q_h$ is in (\ref{quh1}).
The result in (\ref{quh1}) is the origin of the nonanalytic dependence on $y$.

It can now be verified that the term proportional to $\delta_{\nu_n,0}$ in (\ref{rsy3dd}) vanishes, and the equations for $Q_r (i \nu_n)$, $x$, $q_{EA}$ remain unchanged from those in (\ref{ifinaleq1})--(\ref{ifinaleq3}). Indeed, the {\it only\/} change from the $h=0$ solution in Sec.~\ref{sec:0sg} is in the form of $q(u)$ for $x < (q_h/q_{EA}) x$ in (\ref{quh}).

Hence, the gapless condition in  (\ref{gapless}) {\it is\/} now satisfied, and the solutions for $Q_r (i \nu_n)$ and $q_{EA}$ remain unchanged from that for the gapless spectrum in Sec.~\ref{sec:solution}.

\paragraph{Free energy:}

Inserting the solution (\ref{quh}), (\ref{quh1}) into (\ref{rsy5h}), we obtain the extension of (\ref{Fsg0}) to nonzero $h$
\begin{align}
   \mathcal{F}_{sg,h} = \frac{q^{}_{EA} r}{\kappa} + \frac{y^2 q_{EA}^5}{5\kappa}
   + \frac{3 (9 h^{10} y)^{1/3}}{40 \kappa} - \frac{1}{2} \chi_{hb} h^2, \label{Fsgh}
\end{align}
which agrees with (\ref{Fsg0}) at $h=0$.

As the introduction of $h$ does not modify the values of $Q_r (i \nu_n)$ and $q_{EA}$, the free energy $\mathcal{F}_Q$ remains the same as that in Sec.~\ref{sec:freersb}.

\subsubsection{Phase diagram}

To obtain the phase boundary between the replica-symmetric and replica-symmetry-breaking phases, we compare their free energies at leading nontrivial order in $y$.

For the gapped replica-symmetric phase we have the free energy given by (\ref{qeaDelta})--(\ref{fq0})
\begin{align}
\mathcal{F}_{rs} & =  \mathcal{F}_{sg,h} + \mathcal{F}_{Q}^0 = \frac{h^2 \Delta}{2 \kappa} - \frac{2}{3 \kappa^2 \beta} \sum_{|\nu_n| < \Lambda} (\nu_n^2 + \Delta^2)^{3/2} - \frac{\left[ \Delta^2 -r \right]^2}{2 \kappa^2 U}  - \frac{1}{2} \chi_{hb} h^2 \,,  \label{e101}
\end{align}
where the gap $\Delta$ is given by the solution of (\ref{rs2}),

For the gapless replica-symmetry-breaking phase we have from (\ref{Fq0sg}) and (\ref{Fsgh})
\begin{align}
    \mathcal{F}_{rsb} =  \mathcal{F}_{sg,h} + \mathcal{F}_{Q}^0 =  \frac{3 (9 h^{10} y)^{1/3}}{40 \kappa}   - \frac{2}{3 \kappa^2 \beta} \sum_{|\nu_n| < \Lambda} |\nu_n|^{3/2} - \frac{r^2}{2 \kappa^2 U}  - \frac{1}{2} \chi_{hb} h^2\,. \label{e102}
\end{align}

\begin{figure*}
    \centering
    \includegraphics[width=\linewidth]{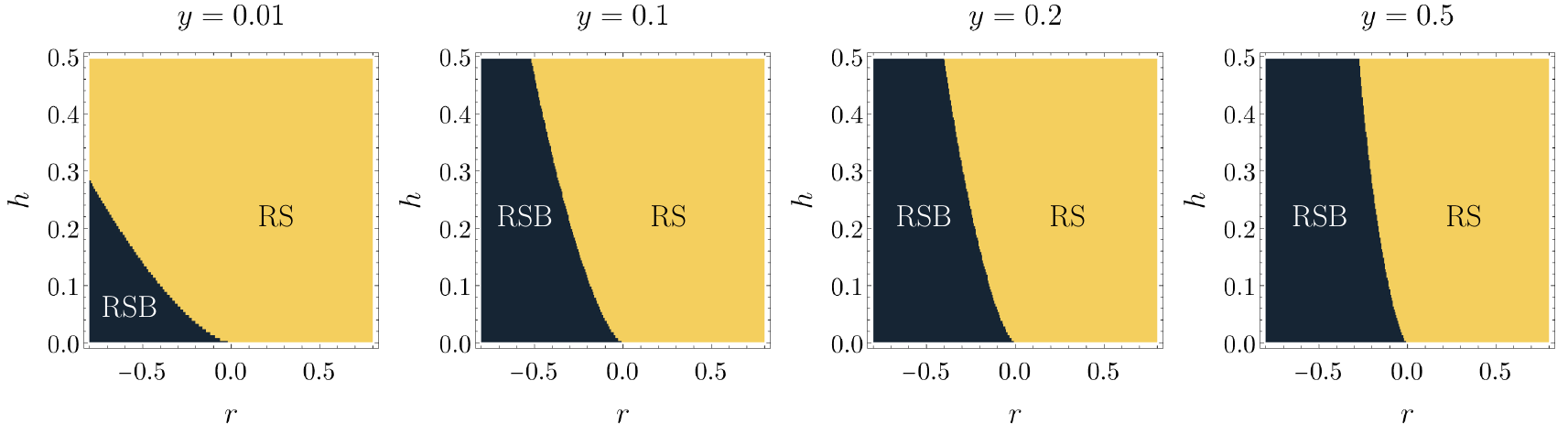}
    \caption{Phase diagram of the Ising model in the $(r,h)$ plane for various values of $y$ at $T=0.2$, illustrating the replica-symmetric (RS) and replica-symmetry-breaking (RSB) phases. The value of $\Delta$ at each point in parameter space is obtained from Fig.~\ref{fig:LTDelta}, and $U, \kappa, \Lambda$ have been set to unity, as previously.}
    \label{fig:LTPD}
\end{figure*}

The summations in (\ref{rs2}), (\ref{e101}), and (\ref{e102}) can be evaluated for $T,\Delta \ll \Lambda$ using identities obtained from (\ref{ebose}): 
\begin{align}
\frac{1}{\beta} \sum_{|\nu_n| < \Lambda} (\nu_n^2 + \Delta^2)^{1/2} & = \frac{\Lambda^2}{2 \pi}
+ \frac{\Delta^2}{4 \pi} \left( 1 + \ln(4\Lambda^2/\Delta^2) \right)
 - 2 \int_\Delta^\infty \frac{d \Omega}{\pi} \frac{(\Omega^2 - \Delta^2)^{1/2}}{e^{\Omega/T} - 1}, \nonumber \\
\frac{1}{\beta} \sum_{|\nu_n| < \Lambda} (\nu_n^2 + \Delta^2)^{3/2} & =  \frac{\Lambda^4}{4 \pi} 
+ \frac{3\Lambda^2 \Delta^2}{4 \pi}
+ \frac{3\Delta^4}{32 \pi} \left( 3 + \ln(16\Lambda^4/\Delta^4) \right)
+ 2 \int_\Delta^\infty \frac{d \Omega}{\pi} \frac{(\Omega^2 - \Delta^2)^{3/2}}{e^{\Omega/T} - 1}.
\end{align}
The phase diagram so obtained in shown in Fig.~\ref{fig:LTPD}; note that due to the $n\rightarrow0$ limit, we have to choose the phase with the maximum free energy \cite{Thouless80}. We observe that the extent of the RSB phase in parameter space, which occurs for $r<0$, shrinks with increasing $h$ and as $y$ is reduced. Note that $\Delta$ remains finite, albeit small, at the transition point for $h > 0$ (it vanishes at the transition for $h=0$), indicating a first-order transition.

\section{Quantum spherical $p$-rotor model}
\label{sec:spherical_model}

We now turn to the analysis of the $p$-rotor model described in Eq.~\eqref{eq:PSM}. The classical infinite-range spin glass with $p$-spin interactions, but without any spherical constraints, was originally introduced by \citet{PhysRevLett.45.79,derrida1981random}. \citet{Gross84} first studied the generalization of this model to nonzero magnetic fields, obtaining an exact solution for $p$\,$\rightarrow$\,$\infty$. The classical spherical model \cite{CC05} can also be solved for any finite $p$, including in the presence of an external field \cite{Crisanti_statics,cavagna1999quenched}. Quantum extensions of these models, by adding a noncommuting transverse field, have been investigated both with (for $p=3$) \cite{Cugliandolo_Grempel,Anous:2021eqj} and without (for $p$\,$\rightarrow$\,$\infty$) \cite{goldschmidt1990solvable} the supplementary spherical constraint. However, the problem of the quantum model in a longitudinal field, which we examine next, has remained unexplored so far.

\subsection{Effective action}

To begin, we derive the effective action for the quantum spherical $p$-rotor model, which will then form the basis for our subsequent saddle-point calculations. In the path integral, the spherical constraint \eqref{eq:Constraint} can be enforced using the exponential representation of the Dirac delta function
 \begin{equation}
	\delta\left(\sum_{i=1}^{N}\sigma^{}_i(\tau)\sigma^{}_i(\tau)-N\right)=\int \mc{D} z \exp\left[i\int_0^\beta d\tau\, z(\tau)\left(\sum_{i=1}^{N}\sigma^{}_i(\tau)\sigma^{}_i(\tau)-N\right)\right],
\end{equation}
which is then inserted into Eq.~\eqref{eq:PSM} at the expense of introducing an auxiliary field $z (\tau)$.

At this stage, since the disorder is quenched, we average the free energy $\log Z[J_{i_1 \ldots i_p}]$ over disorder---using the replica trick---instead of the partition function $Z[J_{i_1 \ldots i_p}]$ itself (which would correspond to the annealed average). The replicated partition function is given by
\begin{alignat}{1}
	\overline{Z^n}&=\int dJ_{i_1\dots i_p} P(J_{i_1\dots i_p})\int \mc{D}\sigma_i^a \,\mc{D}z^a \,\exp\Bigg[i\int_0^\beta d\tau\, z^a(\tau)\big(\sigma^a_i(\tau)\sigma^a_i(\tau)-N\big)\\
 \nonumber
 &-\int_0^\beta d\tau \Bigg(\frac{1}{2g}\dot{\sigma}^a_i(\tau)\dot{\sigma}^a_i(\tau)+\sum_{i_1 <  \ldots < i_p } J_{i_1\dots i_p}\sigma^a_{i_1}(\tau)\dots\sigma^a_{i_p}(\tau) + h\sum_i \sigma_i^a (\tau) \Bigg)
 \Bigg],
\end{alignat}
where $a = 1, \ldots, n$ is the replica index.
Now, we can perform the disorder average by evaluating simple Gaussian integrals to find
\begin{alignat}{1}
 \nonumber
	\overline{Z^n}=\int \mc{D}\sigma_i^a \,\mc{D}z^a \,\exp\Bigg[&-\int_0^\beta d\tau \Bigg(\frac{1}{2g}\dot{\sigma}^a_i(\tau)\dot{\sigma}^a_i(\tau)-i z^a(\tau)\big(\sigma^a_i(\tau)\sigma^a_i(\tau)-N\big) + h\sum_i \sigma_i^a (\tau) \Bigg)\\
 &+\frac{J^2 }{4N^{p-1}}\int_0^\beta \int_0^\beta d\tau\,d\tau'\,\sum_{a,b=1}^n\left(\sum_{i=1}^N\sigma_{i}^a(\tau)\sigma_{i}^b(\tau')\right)^p
 \Bigg].
 \label{eq:dis_avg}
\end{alignat}
Note that the overlap between two different replicas of
the system
\begin{equation}
  Q^{}_{a b}(\tau,\tau')\equiv\frac{1}{N}\sum_{i=1}^N\overline{\langle \sigma_i^a(\tau)\sigma_i^b(\tau')\rangle}, 
\end{equation}
where the overline denotes an average over disorder realizations, appears naturally in Eq.~\eqref{eq:dis_avg}.
We then insert the identity
\begin{align}
1&=\int \mc{D}Q_{ab}\;\delta\left(N\,Q_{ab}(\tau,\tau')-\sum_{i=1}^N 	\sigma_{i}^a(\tau)\sigma_{i}^b(\tau')\right)\nonumber\\
&=\int \mc{D}Q_{ab}\mc{D}\lambda_{ab}\,\exp \left[ i\int_0^\beta\int_0^\beta d\tau d\tau'\, \lambda_{ab}(\tau,\tau')\left(N\,Q_{ab}(\tau,\tau')-\sum_{i=1}^N 	\sigma_{i}^a(\tau)\sigma_{i}^b(\tau')\right)\right] \,\label{eq:deltafunc}
\end{align}
into the partition function, which, in terms of the collective variables $Q_{ab}, \lambda_{ab}$, reads
\begin{alignat}{2}
	\overline{Z^n}&=\int \mc{D}\sigma_i^a \,\mc{D}z^a \,\mc{D}Q_{ab}\,\mc{D}\lambda_{ab}\,\exp\Bigg[&&-\int_0^\beta d\tau \Bigg(\frac{1}{2g}\dot{\sigma}^a_i(\tau)\dot{\sigma}^a_i(\tau)-i z^a(\tau)\sigma^a_i(\tau)\sigma^a_i(\tau) + h\sum_i \sigma_i^a (\tau) \Bigg)\\
\nonumber
 & &&- i \int_0^\beta \int_0^\beta d\tau\,d\tau'\, \sigma_i^a(\tau) \lambda^{}_{ab} (\tau, \tau') \sigma_i^b (\tau')
 \Bigg]\\
\nonumber
 &\times \exp\Bigg[-N i\sum_{a=1}^n\int_0^\beta d\tau   z^a(\tau)+N&&\int_0^\beta \int_0^\beta d\tau d\tau' \left(i\lambda_{ab}(\tau,\tau')Q_{ab}(\tau,\tau')+\frac{J^2 }{4}\sum_{a,b=1}^n \left(Q_{ab}(\tau,\tau')\right)^p\right)\Bigg].
\end{alignat}
Here, we have replaced bilinears of $\sigma_i$ with $Q_{ab}$ wherever possible and collected in the first two lines all the remaining terms involving the $\sigma_i$ fields, which we will integrate out in the next step. Using, for concreteness, the convention
\begin{alignat*}{1}
    \int \mc{D} \vect{v}(x) \exp \left[-\frac{1}{2} \int \int d x\, dx'  \vect{v}(x) \,\mathbf{M} (x, x') \,\vect{v}(x')+ \int dx\, \vect{j}(x) \,\vect{v}(x) \right]\phantom{,} \\
    = \sqrt\frac{(2\pi)^n}{\det \mathbf{M}} \exp \left[\frac{1}{2} \int \int d x\, dx' \vect{j}(x)\, \mathbf{M}^{-1} (x, x') \,\vect{j}(x') \right]
\end{alignat*}
for $n$-dimensional vectors $\vect{v}, \vect{j}$ and an $n\times n$ matrix $\mathbf{M}$, we obtain
\begin{alignat}{2}
	\overline{Z^n}&=\int \mc{D}z^a \,\mc{D}Q_{ab}\,\mc{D}\lambda_{ab}\,\mathrm{det}^{-N/2}\left[-\frac{1}{\pi}\delta_{ab}\delta(\tau-\tau')\left(\frac{1}{2g}\partial_{\tau'}^2+i\, z^a(\tau)\right)+\frac{i}{\pi}\lambda_{ab}(\tau,\tau')\right]\\
\nonumber
 & \times \exp \left\{\frac{N h^2}{2} \int_0^\beta \int_0^\beta d\tau\,d\tau'\, \sum_{ab} \left(2\mathbf{\mc{G}}(\tau,\tau')+ 2 i \mathbf{\lambda}(\tau,\tau')\right)^{-1}_{ab} \right \}\\
\nonumber
 &\times \exp\Bigg[-N i\sum_{a=1}^n\int_0^\beta d\tau   z^a(\tau)+N\int_0^\beta \int_0^\beta d\tau d\tau' \Bigg(i\lambda_{ab}(\tau,\tau')Q_{ab}(\tau,\tau')+\frac{J^2 }{4}\sum_{a,b=1}^n \left(Q^{}_{ab}(\tau,\tau')\right)^p\Bigg)\Bigg],\\
 &\equiv \int \mc{D}z^a \,\mc{D}Q_{ab}\,\mc{D}\lambda_{ab} \exp (- S_{\rm eff}),
\end{alignat}
with the matrix $\mathbf{\mc{G}}$ defined in replica and imaginary-time space as
\begin{equation}
    \mc{G}^{}_{ab} (\tau, \tau') \equiv -\delta_{ab}\,\delta(\tau-\tau')\bigg(\frac{1}{2g}\partial_{\tau'}^2+i\, z^a(\tau)\bigg).
\end{equation}
The replicated effective action can thus be extracted as
\begin{alignat}{1}
\label{eq:Seff}
	\frac{S_{\rm eff}}{N}&=\frac{1}{2}\log \det \left[\frac{1}{\pi} \left(\mc{G}^{}_{ab}(\tau,\tau')+ i \lambda^{}_{ab}(\tau,\tau')\right)\right]+i\sum_{a=1}^n\int_0^\beta d\tau\,  z^a(\tau)\\
 \nonumber
 &-\int_0^\beta \int_0^\beta d\tau\,d\tau'\,\left[i\lambda_{ab}(\tau,\tau')Q_{ab}(\tau,\tau')+\frac{J^2 }{4}\sum_{a,b=1}^n \left(Q^{}_{ab}(\tau,\tau')\right)^p + \frac{h^2}{4} \sum_{a,b=1}^n\left(\mathbf{\mc{G}}(\tau,\tau')+ i \mathbf{\lambda}(\tau,\tau')\right)^{-1}_{ab} \right]~.
\end{alignat}
One of the advantages of the replica approach we are now positioned to harness is that since the effective action \eqref{eq:Seff} is proportional to $N$, the saddle-point approximation is exact in the limit $N\rightarrow\infty$.

\subsection{Saddle-point equations}

We will only be interested in the saddle point of the theory (\ref{eq:Seff}), in which case we can assume that all fields are time-translation invariant and transform to Matsubara frequency space, as in (\ref{rsy3}). Then, we have
\begin{alignat}{1}
\label{eq:Seff2}
	\frac{S_{\rm eff}}{N}&=\frac{1}{2}\sum_{\omega_n} \log\det \left[\frac{1}{\pi} \left( \delta_{ab} \bigg( \frac{1}{2g} \omega_n^2 - i z^a \bigg) +   i \lambda^{}_{ab}(\omega_n)\right)\right]+i \beta \sum_{a=1}^n z^a
\\
 &-\sum_{a,b=1}^n \sum_{\omega_n} \,i\lambda_{ab}(-\omega_n)Q_{ab}(\omega_n) - \frac{\beta h^2}{4} \bigg[ -i \delta_{ab}  z^a  +   i \lambda^{}_{ab}(\omega_n=0) \bigg]^{-1}_{ab}  -\frac{J^2 }{4} \int_0^\beta \int_0^\beta d\tau d\tau'  \left(Q^{}_{ab}(\tau-\tau')\right)^p~  \nonumber.
\end{alignat}
We now employ a slight change of notation to obtain expressions similar to those in Ref.~\onlinecite{QPM}, defining
\begin{align}
- iz^a  = \frac{\overline{\lambda}}{2g} \quad , \quad i \lambda_{ab} (\omega_n) = -\frac{ \Sigma_{ab} (\omega_n)}{2g},
\end{align}
in terms of which, the effective action is
\begin{alignat}{1}
\label{eq:Seff3}
	\frac{S_{\rm eff}}{N}&=\frac{1}{2}\sum_{\omega_n} \log\det \left[\frac{1}{2\pi g} \left( \delta_{ab} \bigg( \omega_n^2 + \overline{\lambda} \bigg)   - \Sigma^{}_{ab}(\omega_n)\right)\right] - \beta \sum_{a=1}^n \frac{\overline{\lambda}}{2g}
\\
 & \nonumber + \sum_{a,b=1}^n \Bigg( \frac{1}{2g}\sum_{\omega_n} \, \Sigma_{ab}(-\omega_n)Q_{ab}(\omega_n) - \frac{\beta h^2 g}{2} \bigg[ \overline{\lambda} \delta_{cd}   -  \Sigma^{}_{cd}(\omega_n=0) \bigg]^{-1}_{ab}  -\frac{J^2 }{4} \int_0^\beta \int_0^\beta d\tau d\tau'  \left(Q^{}_{ab}(\tau-\tau')\right)^p \Bigg) ~.
\end{alignat}
Now, we use the identity (\ref{CSidentity}) to absorb the $h^2$ term into the $\log \det$ as
\begin{alignat}{1}
\label{eq:Seff3z}
	\frac{S_{\rm eff}}{N}&=\frac{1}{2}\sum_{\omega_n} \log\det \left[\frac{1}{2\pi g} \left( \delta_{ab} \bigg( \omega_n^2 + \overline{\lambda} \bigg)   - \Sigma^{}_{ab}(\omega_n) - \beta h^2 g \delta_{\omega_n,0} \right)\right] - \beta \sum_{a=1}^n \frac{\overline{\lambda}}{2g}
 \nonumber\\
 & + \sum_{a,b=1}^n \Bigg( \frac{1}{2g}\sum_{\omega_n} \, \Sigma_{ab}(-\omega_n)Q_{ab}(\omega_n) -\frac{J^2 }{4} \int_0^\beta \int_0^\beta d\tau d\tau'  \left(Q^{}_{ab}(\tau-\tau')\right)^p \Bigg) ~.
\end{alignat}
Then, the saddle-point equation with respect to $ \Sigma_{ab} (\omega_n)$ is
\begin{align}
Q_{ab} (\omega_n) & = g \bigg[ \delta_{ab} \bigg( \omega_n^2 + \overline{\lambda} \bigg)   - \Sigma^{}_{ab}(\omega_n) - \beta h^2 g \delta_{\omega_n,0} \bigg]^{-1}_{ab}.  \label{finaleq2a} 
\end{align}
Inserting this back into (\ref{eq:Seff3z}), we obtain an effective action just for $Q_{ab}$ and $\overline{\lambda}$:
\begin{alignat}{1}
\label{eq:Seff4}
	\frac{S_{\rm eff}}{N}=&-\frac{1}{2}\sum_{\omega_n} \log\det Q_{ab} (\omega_n)  - \beta \sum_{a=1}^n \frac{\overline{\lambda}}{2g}\\
 &
 + \sum_{a,b=1}^n \Bigg( \frac{1}{2g}\sum_{\omega_n} \left[ (\omega_n^2 + \overline{\lambda}) \delta_{ab} - \beta h^2 g \delta_{\omega_n,0} \right] Q_{ab}(\omega_n) 
 \nonumber -\frac{J^2 }{4} \int_0^\beta \int_0^\beta d\tau d\tau'  \left(Q^{}_{ab}(\tau-\tau')\right)^p \Bigg) ~.
\end{alignat}
From this, the saddle-point equations for $Q_{ab}$ and $\overline{\lambda}$ can be read off as
\begin{align}
1 & = \frac{1}{\beta} \sum_{\omega_n} Q_{aa} (\omega_n),  \nonumber \\
\Sigma_{ab} (\tau) & =  \frac{p g J^2}{2}  \left(Q^{}_{ab}(\tau)\right)^{p-1}, \nonumber \\
g Q^{-1}_{ab} (\omega_n) & = \delta_{ab} ( \omega_n^2 + \overline{\lambda}) - \Sigma^{}_{ab}(\omega_n) - \beta h^2 g \delta_{\omega_n,0},  \label{finaleq} 
\end{align}
with the Fourier transform defined by
\begin{align}
    Q_{ab} (\omega_n) = \int_0^{\beta} d \tau\, Q_{ab} (\tau) e^{i \omega_n \tau}\,,
\end{align}
and similarly for $\Sigma_{ab}(\omega_n)$.
We numerically study these equations for the replica-symmetric and one-step replica-symmetry-breaking cases for specific values of $p$. In the discussion below, we focus on the $p$\,$=$\,$3$ model since, as we show below, it exhibits an interesting and nontrivial phase diagram due to the competition between these solutions. 

For completeness, the solution of the $p=2$ model is presented in Appendix \ref{sec:p2spherical}. Note that with $p=2$ and $h=0$, Eq.~(\ref{finaleq}) reduce to (33.32) and (33.33) in \cite{QPM}.
Moreover, for $p$\,$=$\,$4$, $h$\,$=$\,$0$  the equations (\ref{finaleq}) are very similar to Eqs.~(S50--S53) in Ref.~\onlinecite{Christos:2022lma}, the main difference being that $\omega_n^2$ is replaced by $i \omega_n$.

\subsection{Replica-symmetric solution}

In this section, we numerically determine the replica-symmetric (RS) solution of the $p=3$ model as a function of transverse and longitudinal fields ($g$ and $h$, respectively), while choosing the temperature to be close to zero.

The ansatz (\ref{rsy4})  for the replica overlap functions in the RS case necessarily has a replica off-diagonal component 
\begin{align}
    Q_{ab} (\omega_n) &= \beta q \delta_{\omega_n,0} + Q_r (\omega_n) \delta_{ab}, \nonumber \\
  \Sigma_{ab} (\omega_n) &= \beta \varrho \delta_{\omega_n,0} + \Sigma_r (\omega_n) \delta_{ab}, \label{Qansatz}
\end{align}
since, for nonzero $h$, the static moment is always nonzero. Then, using the identities in Appendix~\ref{app:replica}, the equations in (\ref{finaleq}) become (for $p=3$)
\begin{align}
1 & = q + \frac{1}{\beta} \sum_{\omega_n} Q_r (\omega_n), \label{e1a} \\
\Sigma_r (\tau) & =  \frac{3gJ^2}{2} (q + Q_r (\tau))^2 -  \frac{3gJ^2}{2} q^2, \label{e2a} \\
\varrho  & =  \frac{3}{2}gJ^2 q^2, \label{e3a} \\
    \frac{g}{Q_r (\omega_n)} & = \omega_n^2 + \overline{\lambda} - \Sigma_r (\omega_n),  \label{e4a} \\
      \frac{g\, q}{Q_r(\omega_n = 0)^2} & = \varrho +  g h^2.  \label{e5a} 
\end{align}

\begin{figure}
\center{\includegraphics[width=3.15in]{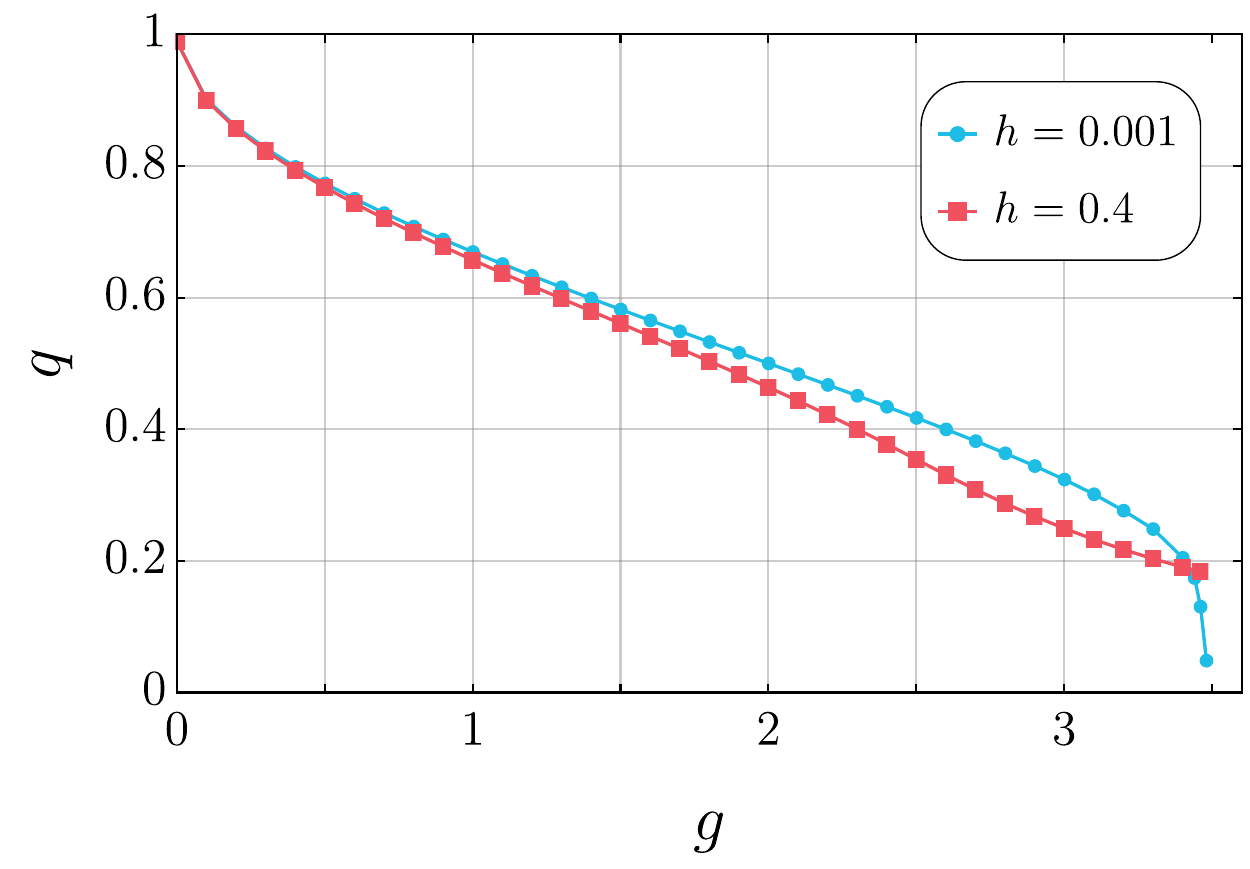}}
\caption{Behavior of the order parameter $q$ for the replica-symmetric solution of the quantum spherical $p=3$ model as a function of the transverse field $g$ for different values of longitudinal fields $h$ (with $J=1$, $T=0.005$). The phase transition between the replica-symmetric and paramagnetic solutions occurs only for $h=0$ and is continuous. }
\label{fig:quan_p3_RS}
\end{figure}

We solve these equations self-consistently in the imaginary-frequency domain for the order parameter $q$ as a function of the transverse field $g$ at low temperatures for several values of the longitudinal field, as shown in Fig.~\ref{fig:quan_p3_RS}. In this model, the phase transition between the RS phase ($q\neq 0$) and the paramagnet (PM, $q=0$) occurs only at zero longitudinal field. With even a slight increase in $h$, the replica-symmetric solution becomes equivalent to the paramagnet for all values of $g$, as in Sec.~\ref{sec:hrs}. 

Regardless of the value of the longitudinal field, we can show that the spectrum of the RS solution is gapped. To do so, we compute the dynamic spin susceptibility of the system, which requires analytic continuation of the equations \eqref{e1a}--\eqref{e5a} to real frequencies. Taking the zero-temperature limit, we obtain equations of the form
\begin{align}
\rho(\omega) &=\frac{g}{\pi} \frac{\Sigma_r''(\omega)}{(\Sigma_r''(\omega))^2+(-\omega^2+\olambda-\Sigma_r'(\omega))^2},\label{eq:RS_real_T0_1}\\
\Sigma_r''(\omega) &= 3\pi gJ^2 \left(q \rho(\omega) + \frac12\int_{0}^{\omega}d\omega_1\rho(\omega_1)\rho(\omega-\omega_1)\right),\\
\Sigma_r'(\omega)&=2\int_{0}^{+\infty} \frac{d\nu}{\pi}\frac{\nu\Sigma_r''(\nu)-\omega\Sigma_r''(\omega)}{\nu^2-\omega^2},\\
\varrho  & =  \frac{3}{2} gJ^2q^{2},\\
q&=\frac{g\varrho+g^2h^2}{[\overline{\lambda} - \Sigma_r'(\omega=0)]^2},\\
1&=q+\int_0^{+\infty}d\omega \rho(\omega),\label{eq:RS_real_T0_2}
\end{align}
where the spectral representation of the regular component of the replica overlap $Q_{ab}(i\omega_n)$ is
\begin{align}
Q_r(z)=\int_{-\infty}^{\infty} d\omega\frac{\omega\, \rho(\omega)}{z^2+\omega^2},
\end{align}
and $\Sigma_r'(\omega)$ and $\Sigma_r''(\omega)$ are the real and imaginary parts of $\Sigma_{r}(i\omega_n)$, respectively.
The spectral function is related to the retarded Green's function as $\rho(\omega)= \text{Im}\,Q_r(\omega)/\pi$. Taking the $z\to \infty$ limit above, the associated sum rule becomes
$\int_{0}^{+\infty}d\omega\, \omega \rho(\omega) ={g}/2$.

We note that the equations above, strictly speaking, permit several solutions, one of which has a discontinuity at zero frequency given by $\rho(\omega=0)=1/({2\pi J \sqrt{6q}})$. To avoid this, we initialize the system with a simple step function with a small gap and let the equations converge (defined as when the error reaches $10^{-20}$). As a final step, we also check that the sum rule is satisfied. 

\begin{figure}
(a)\includegraphics[width=3.15in]{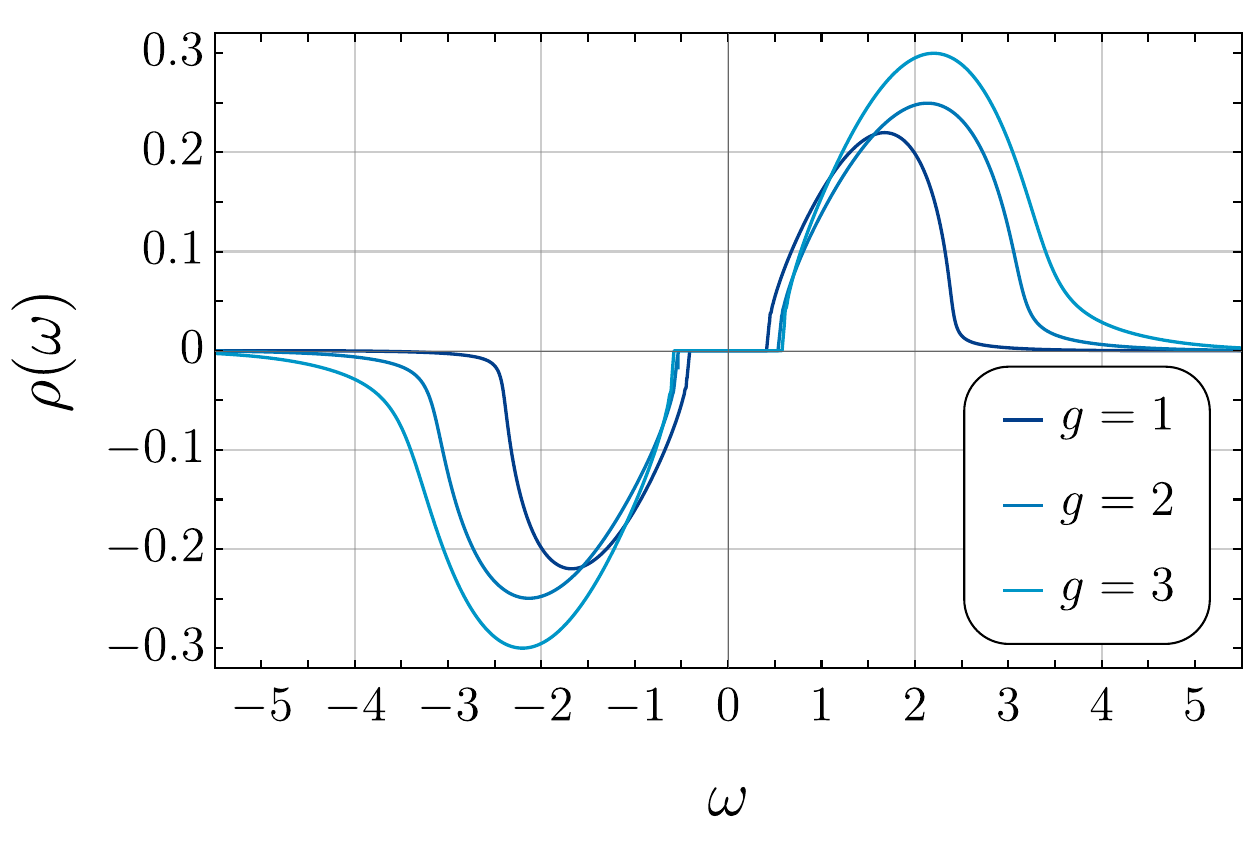}\,\,\,\,\,
(b)\includegraphics[width=3.15in]{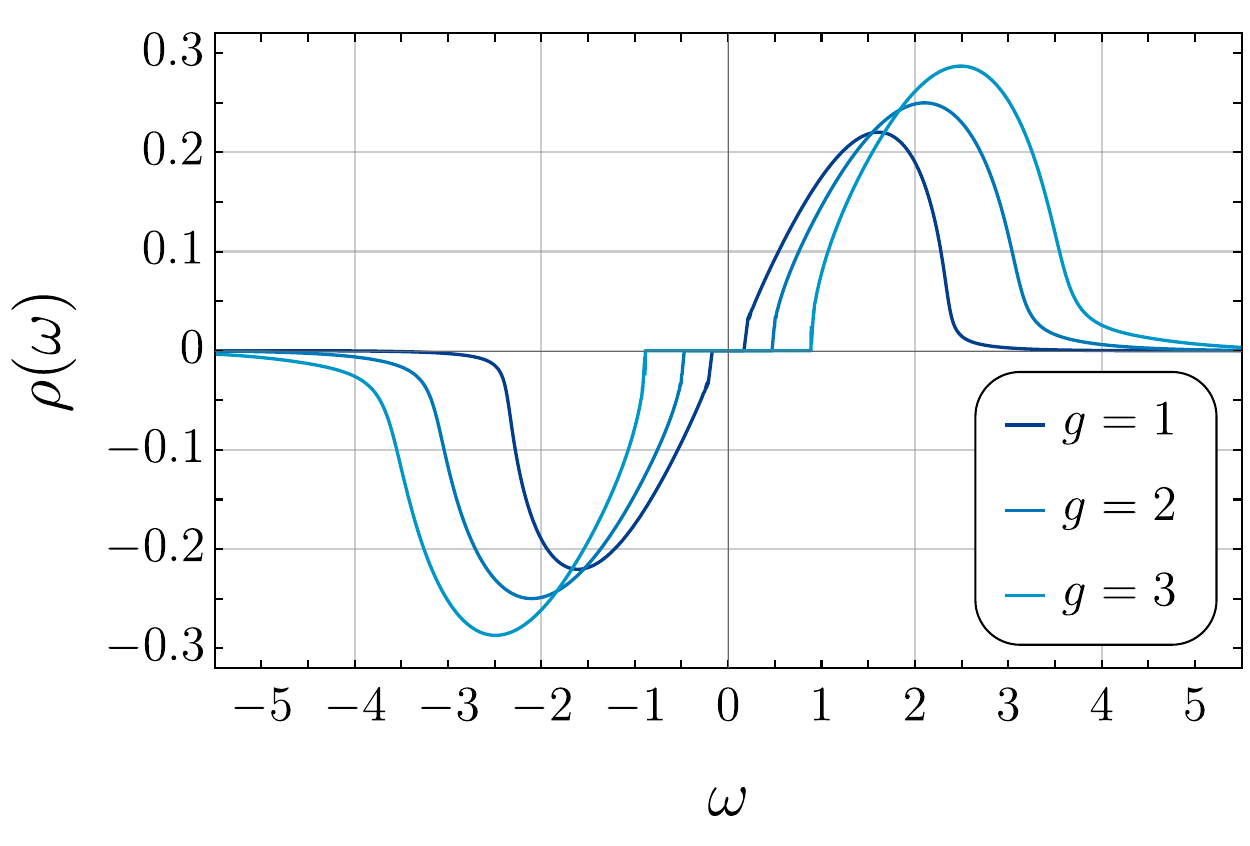}
\caption{Spectral functions of the quantum spherical $p=3$ model for (a) $h=0$ and (b) $h=1$ at zero temperature (with $J=1$). (a) The phase transition between the RS phase and the quantum paramagnet occurs at a finite $g>3$, and the gap thus eventually saturates upon increasing $g$. (b) The replica-symmetric solution persists for all values of $g$ with a gap that grows with increasing $g$.}
\label{fig:rho_T0_h0_g}
\end{figure}

The spectral functions obtained in this fashion are presented in Fig.~\ref{fig:rho_T0_h0_g}. For $h=0$, we observe that as $g$ increases, the gap saturates to a finite value and eventually, after the transition point, the system becomes a quantum-disordered paramagnet, as shown in Appendix~\ref{sec:largeg}. However, at a finite field, the gap increases with increasing $g$ without a visible saturation since the replica-symmetric solution and the paramagnet are one and the same.

\subsection{One-step replica symmetry breaking}
In this section, we now allow for nontrivial structure in the replica off-diagonal space. In general, replacing Eq.~(\ref{Qansatz}), we can write
\begin{align}
    Q_{ab} (\omega_n) & = \beta q_{ab} \delta_{\omega_n, 0}; \quad a \neq b, \nonumber \\
    \Sigma_{ab} (\omega_n) & = \beta \varrho_{ab} \delta_{\omega_n, 0}; \quad a \neq b. 
\end{align}
The matrices $q_{ab}$ and $\varrho_{ab}$ are characterized by Parisi functions $q(u)$ and $\varrho (u)$ with $u$\,$\in$\,$[0,1]$. For the diagonal components, we make the same ansatz as in (\ref{Qansatz})
\begin{align}
    Q_{aa} (\omega_n) &= \beta q_1 \,\delta_{\omega_n,0} + Q_r (\omega_n),  \nonumber \\
  \Sigma_{aa} (\omega_n) &= \beta \varrho_1 \, \delta_{\omega_n,0} + \Sigma_r (\omega_n).  \label{Qansatz2}
\end{align}
We consider the one-step replica-symmetry-breaking (RSB) ansatz, for which (see Fig.~\ref{fig:qup})
\begin{equation}\label{q1_rho1_rsb}
q (u) = 
    \begin{cases}
        q_1 \,, \quad \mbox{$x < u < 1$}\\
        q_0 \,, \quad \mbox{$0 < u < x$}
    \end{cases},
    \quad \mbox{and}\quad
    \varrho(u) = 
    \begin{cases}
        \varrho_1 \,, \quad\mbox{$x<u<1$}\\
        \varrho_0 \,, \quad \mbox{$0 < u < x$} 
    \end{cases}.
\end{equation}
Using the identities in Appendix~\ref{app:replica}, the equations in (\ref{finaleq}) now become
\begin{align}
1 & = q_1 + \frac{1}{\beta} \sum_{\omega_n} Q_r (\omega_n), \label{e1b} \\
\varrho_1 + \Sigma_r (\tau) & =  \frac{pgJ^2}{2} (q_1 + Q_r (\tau))^{p-1}, \label{e2b} \\
\varrho_1 & =  \frac{pgJ^2}{2} q_1^{p-1}, \label{e3b} \\
 \varrho_0 & =  \frac{pgJ^2}{2} q_0^{p-1}, \label{e4b} \\
  Q_r (\omega_n) & = \frac{g}{\omega_n^2 + m^2 - \Sigma_r (\omega_n) + \Sigma_r (\omega_n = 0)},\label{e5bb} \\
        g h^2   +  \varrho_1 &= \frac{q_1 g + m^2 \beta x(q_1-q_0)^2 }{ [g/m^2 + \beta x (q_1 - q_0)]^2}, \label{e6bb} \\ 
       g h^2   +  \varrho_0 &= \frac{g q_0}{[g/m^2 + \beta x (q_1 - q_0)]^2}, \label{e7bb} 
\end{align}
where we have simplified the equations by changing variables and introducing $m^2 = \overline{\lambda} - \Sigma_r (\omega_n = 0)$.

There are eight unknowns, namely, $m^2$, $Q_r$, $\Sigma_r$, $q_1$, $q_0$, $\varrho_1$, $\varrho_0$, and $\beta x$, in the equations above. All but one of these variables can be determined by solving the seven equations (\ref{e1b})--(\ref{e7bb}). However, the value of the breakpoint $\beta x$ is undetermined and can be fixed either by demanding a gapless solution, as in Appendix 3 of Ref.~\onlinecite{Christos:2022lma}, or by requiring the free energy to be stationary with respect to $x$. The additional equation then makes the system complete and the solution can be obtained by solving the set of equations self-consistently. 
In the following, we derive this extra equation in both cases: by imposing the gapless constraint, and by using the free energy.

\subsubsection{Gapless condition}
Let us first determine the condition for a gapless spectrum in the RSB case; our approach here is similar to the analysis in Ref.~\onlinecite{Christos:2022lma}.
For a gapless spectrum, we expect the low-frequency expansion of $Q_r$, for real $\omega$, to have the form
 \be
Q_r (\omega + i \eta) = \frac{g}{m^2} + |\omega|^{\alpha} \left[ a + i b\, \mbox{sgn}(\omega) \right] + \ldots, \label{gps1}
\ee
for some $\alpha$\,$>$\,$0$, $b$\,$>$\,$0$. From (\ref{e2b}), we find that the leading singularity in $\Sigma_r(\omega)$ is given by the linear-in-$Q_r(\omega)$ term in (\ref{e2b}), so
\bea
\mbox{Im}\, \Sigma_r (\omega + i \eta) & = & \frac{p(p-1) g J^2}{2} q_1^{p-2} \mbox{Im}\, Q_r (\omega + i \eta), \nonumber \\ &=& \frac{p(p-1) g J^2}{2} q_1^{p-2} b \,|\omega|^{\alpha}\, \mbox{sgn} (\omega)\,. \label{gps2}
\eea
However, from (\ref{e5bb}), for low frequencies, we also have
\be
\mbox{Im}\, \Sigma_r (\omega + i \eta) = - g\, \mbox{Im} \left[ Q_r (\omega + i \eta) \right]^{-1} =  \frac{m^4}{g} b \, |\omega|^\alpha \, \mbox{sgn} (\omega)\,.\label{gps3}
\ee
Therefore, comparing equations (\ref{gps2}) and (\ref{gps3}), we arrive at the constraint on the variable $m$ that has to be satisfied for the gapless solution to exist:
\be
m^4 = \frac{p(p-1) g^2 J^2}{2} q_1^{p-2}\,. \label{gps4}
\ee
It is easy to check that for the special case of $p=2$, this expression precisely agrees with Eq.~(33.41) in Ref.~\onlinecite{QPM} and is similar to Eq.~(4.11) in Ref.~\onlinecite{Anous:2021eqj}.

It is important to note that upon solving the replica-symmetric equations self-consistently, we did not obtain a gapless solution. One way to understand this is to plug the gapless constraint \eqref{gps4} into the equations for the replica-symmetric case \eqref{e1a}--\eqref{e5a}, which expressly demonstrates that the constraint cannot be satisfied for any values of $h$. Therefore, the replica-symmetric solution always has a gap.

\subsubsection{Free energy}

Another way to fix $x$ is to determine the saddle point of the free energy with respect to $x$. To do so, we rewrite the effective action of the $p$-rotor model (\ref{eq:Seff4}) explicitly for the one-step replica-symmetry-breaking case
\begin{alignat}{1}
\label{eq:Seff40}
	\frac{S_{\rm eff}}{Nn}= &-\frac{1}{2} \sum_{\omega_n} \log[Q_r (\omega_n)] 
 - \frac{1}{2} \frac{\beta q_0}{Q_r (\omega_n = 0) + \beta x (q_1 - q_0)} + \frac{1}{2x} \ln \frac{ Q_r (\omega_n = 0)}{Q_r (\omega_n = 0) + \beta x (q_1 - q_0)} \nonumber \\
 &
 -  \frac{\beta \overline{\lambda}}{2g}
  + \frac{1}{2 g} \sum_{\omega_n} (\omega_n^2 + \overline{\lambda} ) Q_r (\omega_n) 
 + \frac{\overline{\lambda} \beta q_1}{2g} - \frac{\beta h^2}{2}\left[Q_r (\omega_n = 0) + \beta x(q_1 - q_0) 
 \right] \nonumber \\
 & -\frac{\beta J^2 }{4}  \int_0^\beta d\tau  \left[ \left(Q_r (\tau) + q_1\right)^p - q_1^p (1-x) - q_0^p x \right]. 
 \end{alignat}
This action allows one to directly verify that the saddle-point equations are in fact correct: taking the derivative of (\ref{eq:Seff40}) with respect to $q_0$ and $\overline{\lambda}$, we see that the equations (\ref{e1b}) and (\ref{e7bb}) indeed hold after a change of variables to $m^2 = \overline{\lambda} - \Sigma_r (\omega_n = 0)$. 

To obtain an equation for $x$, we differentiate the action with respect to the breakpoint and obtain
\begin{align}
    \frac{\partial S_{\rm eff}}{\partial x} = 0 
     \quad \Rightarrow \quad &
     \frac{J^2 \beta^2}{4} (q_1^p - q_0^p) - \frac{1}{2x^2}\log \left[ 1 + \frac{\beta x (q_1-q_0)}{ Q_r (\omega_n = 0)}\right] \nonumber \\ & + \frac{\beta^2 h^2}{2} (q_1 - q_0) + \frac{ \beta (q_1 - q_0)(Q_r (\omega_n =0) + \beta q_1 x - 2 \beta q_0 x)}{2x (Q_r (\omega_n=0) + \beta x (q_1 - q_0))^2} = 0.
     \label{dSdx}
 \end{align}
Although such a saddle-point condition has been employed to obtain the equilibrium state for the classical model \cite{Talagrand}, 
we will not use (\ref{dSdx}) here. 
It what follows, we instead use the gapless constraint to determine the breakpoint $x$. 

\subsubsection{Spectral functions}
As for the replica-symmetric case before, we are interested in the behavior of the spectral functions in the RSB case. To this end, we analytically continue the equations on the imaginary-frequency axis \eqref{e1b}--\eqref{e4b} and \eqref{e5bb}--\eqref{e7bb}, focusing on the $p=3$ case at zero temperature. After such analytic continuation, we obtain
\begin{align}
1&=q_1+\int_0^{+\infty}d\omega \rho(\omega),\\
\Sigma_r''(\omega) &= 3\pi gJ^2 \left(q_1 \rho(\omega) + \frac12\int_{0}^{\omega}d\omega_1\rho(\omega_1)\rho(\omega-\omega_1)\right),\label{eq:sp_dens_RSB1}\\
\Sigma_r'(\omega)&=2\int_{0}^{+\infty} \frac{d\nu}{\pi}\frac{\nu\Sigma_r''(\nu)-\omega\Sigma_r''(\omega)}{\nu^2-\omega^2},\\
\varrho_1 & =  \frac{3gJ^2}{2} q_1^2, \\
 \varrho_0 & =  \frac{3gJ^2}{2} q_0^2,\\
  \rho(\omega) &= \frac{g}{\pi}\frac{\Sigma_r''(\omega)-\Sigma_r''(\omega=0)}{(\Sigma_r''(\omega)-\Sigma_r''(\omega=0))^2+(\omega^2-m^2+\Sigma_r'(\omega)-\Sigma_r'(\omega=0))^2}, \\\label{eq:x}
   g h^2   +  \varrho_1 &=\frac{q_1 g + m^2 \beta x(q_1-q_0)^2 }{ [g/m^2 + \beta x (q_1 - q_0)]^2}, \\ \label{eq:q0}
   g h^2   +  \varrho_0&=\frac{g q_0}{[g/m^2 + \beta x (q_1 - q_0)]^2}.    
\end{align}
For the final equation, as mentioned, we use the gapless constraint
\be
m^4 = 3 g^2 J^2 q_1,\,\label{eq:sp_dens_RSB2}
\ee
derived above in (\ref{gps4}), and this gapless behavior should also be reflected in the spectral functions. Therefore, we are allowed to make a linear ansatz 
\be
\rho(\omega)=\omega\, f(\omega).
\ee
We insert this ansatz in Eqs.~\eqref{eq:sp_dens_RSB1}--\eqref{eq:q0} and solve the equations self-consistently for $f(\omega)$.
It is quite challenging to obtain the self-consistent solution to these equations for all unknowns. Instead, we simplify this task by solving the equations on the imaginary-frequency axis and self-consistently obtain the value of the order parameter $q_1$ at a temperature close to zero with a high precision. Then, we use this result to obtain the behavior of the spectral functions. To ensure that the converged solution is meaningful, we check the sum rule, which, in this case, is given by $\int_{0}^{+\infty}d\omega\, \omega^2 f(\omega) ={g}/2$. 

We sketch the behavior of the spectral functions Im\,$Q_r(\omega)$ in Fig.~\ref{fig:sp_dens_RSB}. From the equations above, it is clear that spectral functions are independent of $h$, $\beta x$ and $q_0$.  Only two parameters, namely, $q_0$ and $\beta x$, depend on $h$. This is similar to the behavior of the Ising model in Sec.~\ref{sec:rsb:hh}, where the
only change from the $h=0$ solution in Sec.~\ref{sec:0sg} was in the form of $q(u)$ for $u < (q_h/q_{EA}) x$ in (\ref{quh}).
It is also instructive to determine the dependence of $q_0$ and $\beta x$ on $h$ in the $h\to0$ limit, which can be calculated from  the equations \eqref{eq:x} and \eqref{eq:q0}. We find that, to the lowest order, the dependence is simply 
\begin{align}
q^{}_0  = \frac{4g^2}{m^4} h^2;\,\,\,\beta x = \frac{g}{q^{}_1 m^2}-\frac{4g^2}{3m^2J^2q_1^3}h^2, 
\end{align}
which, as noted in Sec.~\ref{sec:main_results}, is in distinction to the $h^{2/3}$ scaling observed for the Ising spin glass.

\begin{figure}[tb]
\center{
(a)\includegraphics[width=3.19in]{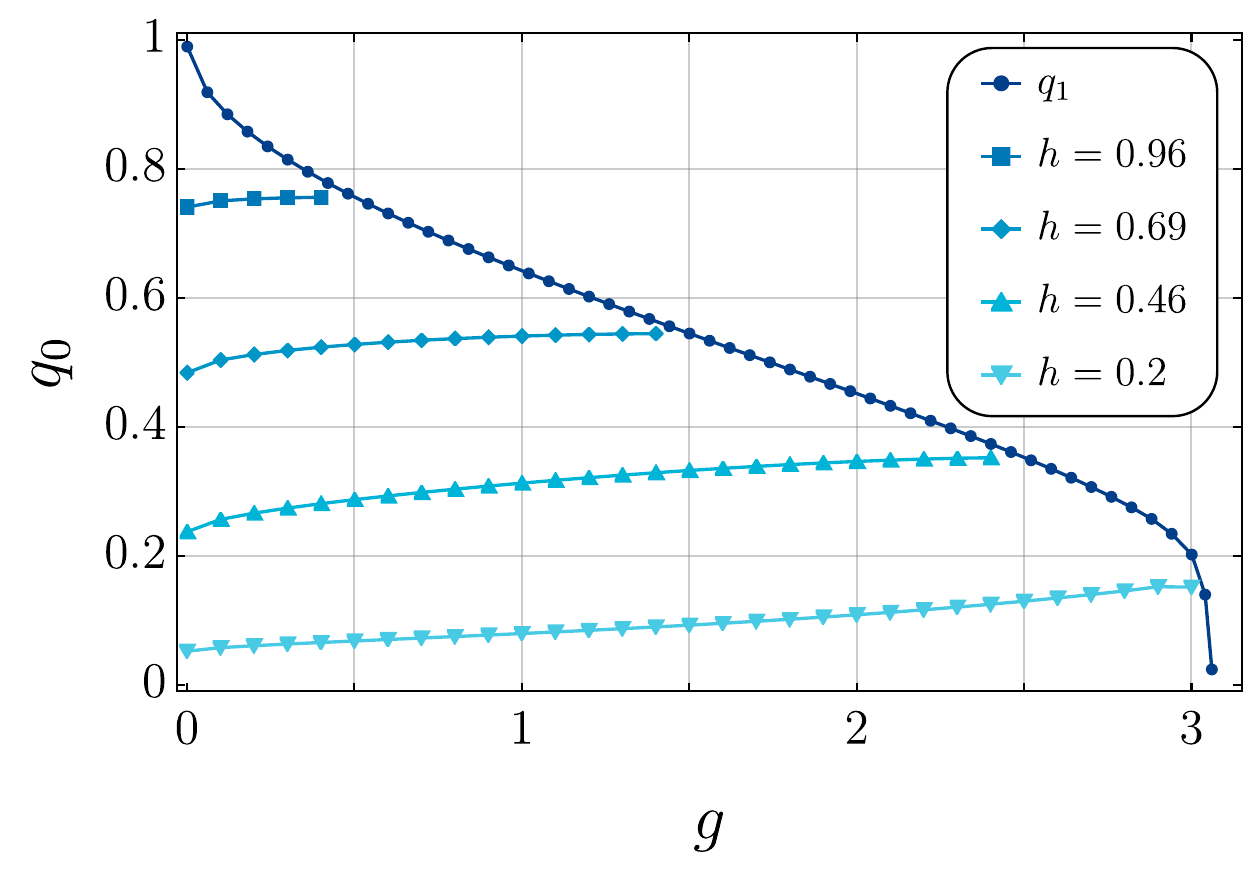}}
(b)\includegraphics[width=3.19in]{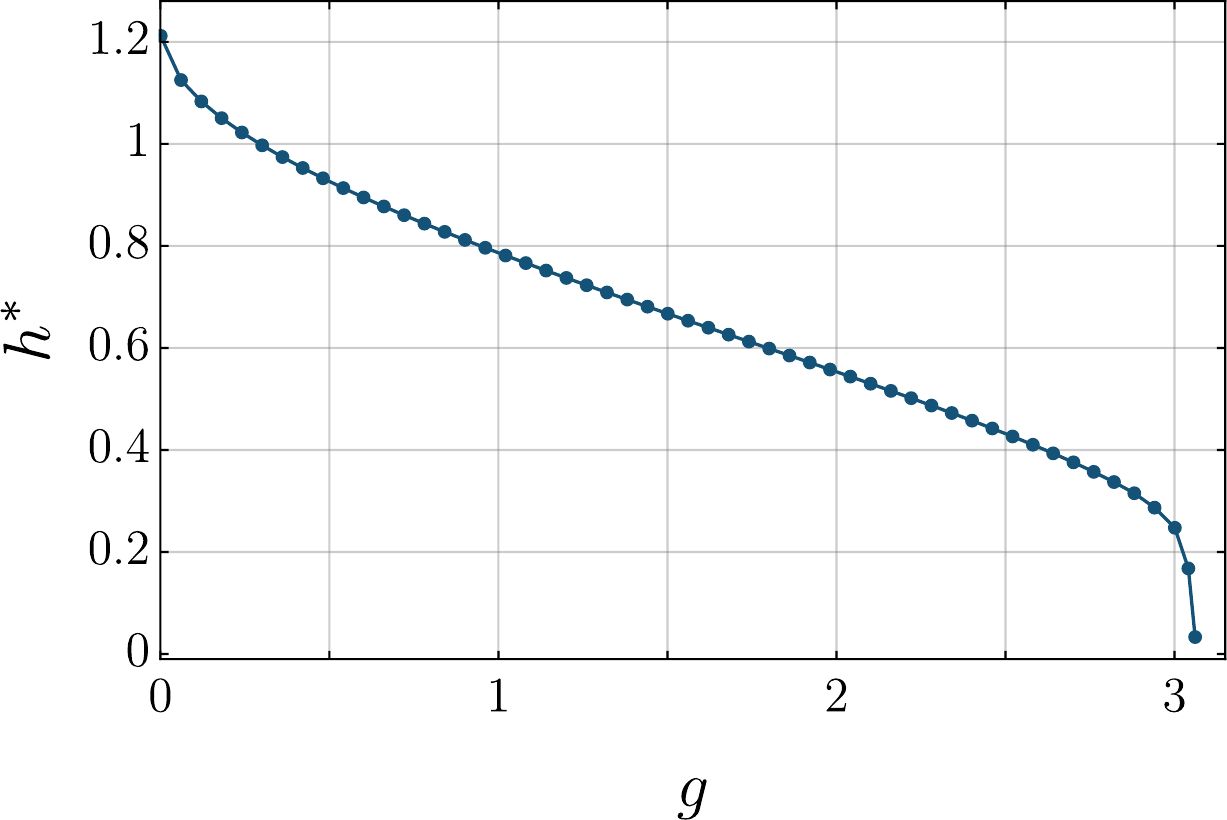}
\caption{(a) Behavior of $q_0$ and $q_1$ as a function of $g$ for the spherical quantum $p$\,$=$\,$3$-rotor model in a longitudinal field with $T=0.005$ and $J=1$. Note that $q_1$ is independent of $h$ but $q_0$ is not. (b) Boundary of stability of the RSB phase as determined by setting $q_0 (g,h^*)$\,$=$\,$q_1 (g)$. For $h>h^*(g)$, the necessary condition $q_0 \le q_1$ required for the RSB solution no longer holds.}
\label{fig:phase_diag_RSB}
\end{figure}

\subsection{Phase diagram}
Finally, having assembled all the necessary ingredients, we now compute the phase diagram of the quantum spherical  $p$\,$=$\,$3$-rotor model in a longitudinal field at low temperatures. To do so, we first determine the boundary of stability of the RSB solution by self-consistently computing the spin glass order parameter from equations \eqref{e1b}--\eqref{e7bb} and comparing $q_0$ and $q_1$. For the stability of the RSB solution, one has to require $q_0$\,$\le$\,$q_1$ \cite{CC05}, and the solution ceases to be physical when $q_0$\,$>$\,$q_1$. Our numerical results in this regard are displayed in Fig.~\ref{fig:phase_diag_RSB}. In particular, Fig.~\ref{fig:phase_diag_RSB}(a) presents $q_0$ and $q_1$ as a function of $g$ for various values of $h$; the RSB solution exists only for $q_0$ that lie below the line $q_1(g)$. Based on this information, Fig.~\ref{fig:phase_diag_RSB}(b) traces, in $g$-$h$ space, the line $h^*(g)$ along which $q_0 (g,h^*)$\,$=$\,$q_1 (g)$. Since this marks the boundary of stability of the RSB solution, the phase transition between the RSB and RS phases may occur at this line.

To check whether this is indeed the case, we compute the free energies of both the RS and RSB solutions.  The free energy in the RS case can be written using \eqref{eq:Seff4} and we obtain  
\begin{alignat}{1}
	\frac{F^{\textsc{rs}}_{\rm eff}}{Nn}=&-\frac{1}{2 \beta}\sum_{\omega_n} \log [(\omega_n^2 + \Gamma^2)Q_r(\omega_n)/g] -\frac{1}{2}\frac{q}{Q_r(\omega_n=0)} -   \frac{\overline{\lambda}}{2g}
 +  \frac{\overline{\lambda} q}{2g} +  \frac{1}{2g \beta}\sum_{\omega_n}  \left[(\omega_n^2 + \overline{\lambda})Q_{r}(\omega_n) - g \right] 
 \nonumber \\
 &- \frac{ h^2 }{2}  Q_{r}(\omega_n=0) -\frac{J^2 }{4}  \int_0^\beta d\tau \left[ \left(q+Q^{}_{r}(\tau)\right)^3 - q^3 \right] + \frac{\Gamma}{2} + \frac{1}{\beta} \ln \left[ 1 - e^{-\beta \Gamma} \right],\label{eq:free_en_RS}
\end{alignat}
where we inserted a factor of $(\omega_n^2 + \Gamma^2)/g$ into the argument of the logarithm to regularize the behavior at large $|\omega_n|$, and subtracted a constant to regulate the frequency summation in the term linear in $Q_r$. This is equivalent to normalizing the path integral with an oscillator of frequency $\Gamma$. In order to make the resulting expression independent of $\Gamma$, we subtracted the free energy of this oscillator. 
For the RSB case, we use the expression for the free energy computed above, focusing on $p=3$, 
\begin{alignat}{1}
	\frac{F^{\textsc{rsb}}_{\rm eff}}{Nn}= &-\frac{1}{2 \beta} \sum_{\omega_n} \log[(\omega_n^2 + \Gamma^2) Q_r (\omega_n)/g] 
 - \frac{1}{2} \frac{q_0}{Q_r (\omega_n = 0) + \beta x (q_1 - q_0)} + \frac{1}{2 \beta x} \ln \frac{ Q_r (\omega_n = 0)}{Q_r (\omega_n = 0) + \beta x (q_1 - q_0)} \nonumber \\
 &
 -  \frac{\overline{\lambda}}{2g}
  + \frac{1}{2 g \beta} \sum_{\omega_n} \left[ (\omega_n^2 + \overline{\lambda} ) Q_r (\omega_n) - g \right]
 + \frac{\overline{\lambda} q_1}{2g} - \frac{ h^2}{2}\left[Q_r (\omega_n = 0) + \beta x(q_1 - q_0) 
 \right] \nonumber \\
 &-\frac{ J^2 }{4}  \int_0^\beta d\tau  \left[ \left(q_1+Q_r (\tau) \right)^3 - q_1^3 (1-x) - q_0^3 x \right] + \frac{\Gamma}{2} + \frac{1}{\beta} \ln \left[ 1 - e^{-\beta \Gamma} \right].\label{eq:free_en_RSB}
 \end{alignat}
Here, we use the same regularization as for the free energy of the replica-symmetric solution. In both cases, $\Gamma$ has to be much smaller than the Matsubara frequency cutoff.

\begin{figure}[tb]
\center{
\includegraphics[width=0.4\linewidth]{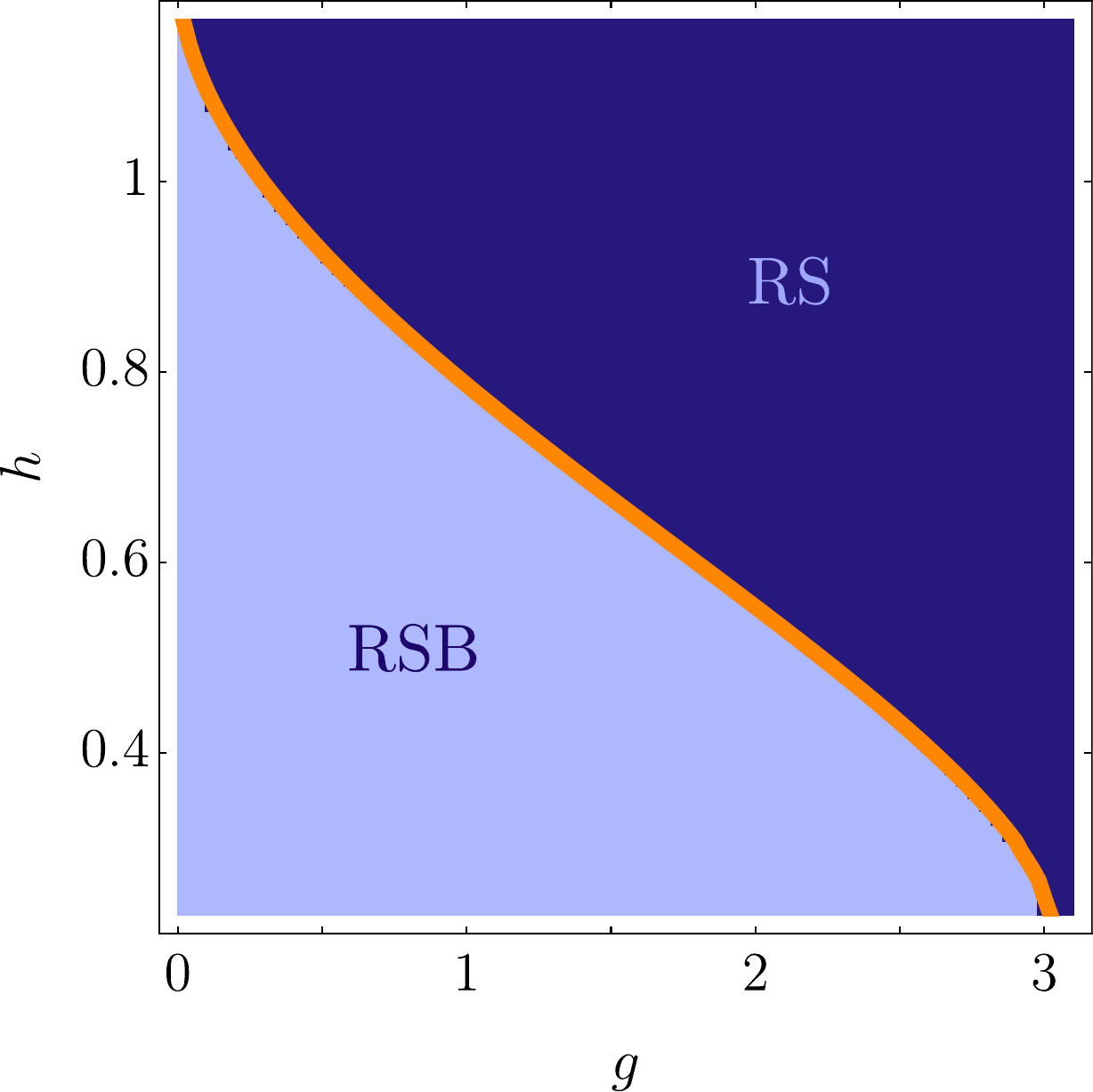}}
\caption{Phase diagram of the spherical quantum $p$\,$=$\,$3$-rotor model in a longitudinal field with $T=0.005$ and $J=1$. The phase diagram is evaluated by calculating the difference between the free energies of the replica-symmetric (RS, \ref{eq:free_en_RS}) and replica-symmetry-breaking (RSB, \ref{eq:free_en_RSB}) solutions. The orange line on the phase boundary traces the function $h^*(g)$ plotted in Fig.~\ref{fig:phase_diag_RSB}(b). }
\label{fig:phase_diag_pspin}
\end{figure}

We evaluate the difference in free energies between the RSB and RS solutions to find the parameter regions where each phase is dominant. Following the discussion in Ref.~\onlinecite{Thouless80}, we pick the ground state as the one that has a larger free energy. The resulting phase diagram is shown in Fig.~\ref{fig:phase_diag_pspin}; we find that the phase boundary obtained by comparing the free energies coincides (to a precision of $\lvert F^{\textsc{rsb}}_{\rm eff}- F^{\textsc{rs}}_{\rm eff}\rvert /({Nn}) < 10^{-5}$) with the line $h^*(g)$ demarcating the stability of the RSB solution in Fig.~\ref{fig:phase_diag_RSB}(b).

\section{Conclusions}

Motivated by the quantum simulation of spatially disordered long-ranged interacting systems for solving combinatorial optimization problems \cite{MIS22,tasseff2022emerging}, we have studied the equilibrium dynamic properties of two quantum spin glass models with infinite-range interactions: the Ising model in 
(\ref{H_Ising}), and the spherical $p$-rotor model in (\ref{eq:PSM}) and (\ref{eq:Constraint}). In addition to random couplings between spins/rotors, both models have an applied longitudinal field $h$, and do not break any global symmetry in any phase. A coupling $g$ tunes the strength of the quantum fluctuations, and at large $g$, we obtain a gapped paramagnetic phase. Our primary interest was in the small-$g$ regime, where replica symmetry breaking leads to a quantum spin glass phase.

For the Ising model, we employed the Landau theory approach of Ref.~\onlinecite{RSY95} to show that the spin glass phase has full replica symmetry breaking, with the Parisi function sketched in Fig.~\ref{fig:qu2}. We computed the spin autocorrelation function in this spin glass phase, and showed that it is generically gapless, with a spectral density which vanishes linearly with frequency (see (\ref{Qrlow}) and Secs.~\ref{sec:solution} and~\ref{sec:rsb:hh}). 

In contrast, the spherical $p$-rotor model ($p \geq 3$) has a different behavior in the spin glass phase. There is only one-step replica symmetry breaking (see Fig.~\ref{fig:qup}), and the breakpoint $x$ remains undetermined by the saddle-point equations for the matrix spin glass order parameter. Rigorous mathematical work on the classical $g=0$ model \cite{Talagrand} has shown that the equilibrium value of $x$ is determined by the stationarity of the free energy with respect to $x$. For this $x$, we showed that the spin glass state has an energy gap. In contrast, if a marginal stability condition \cite{Cugliandolo_Kurchan,Cugliandolo_Lozano,Cugliandolo_Grempel} is used to determine $x$, we found a gapless spectrum with a linear frequency dependence at small frequency (see Fig.~\ref{fig:sp_dens_RSB}), similar to that of the Ising model.

Our results on the nature of the spin glass phase also find relevance in the context of quantum optimization problems since the properties of the gap---or lack thereof---directly inform the feasibility of preparing low-energy states via adiabatic algorithms \cite{RevModPhys.90.015002,Cain23,schiffer2023circumventing}. 

Looking ahead, it would be useful to obtain the results described here without the replica method, using a path integral on the Schwinger-Keldysh contour. This has already been studied for the spherical $p$-rotor model \cite{Cugliandolo_Lozano,Biroli01,Biroli02}, where the breakpoint of one-step replica symmetry breaking, $x$, has been connected to an effective temperature of the aging dynamics. Moreover, the equation for a gapless spectrum in (\ref{gps4}) is the same as Eq.~(6.24) obtained in Ref.~\onlinecite{Cugliandolo_Lozano} from aging dynamics. However, the aging dynamics of the quantum Ising model have not been studied so far, and it would be interesting to  relate that to the full replica symmetry breaking characterized by Fig.~\ref{fig:qu2}. Such a computation would be the analog of earlier studies \cite{Sompolinsky1,Sompolinsky2,Sompolinsky3,Kurchan94,Biroli20} of aging dynamics in classical spin glasses described by Langevin equations, and we expect that there would be significant differences \cite{Parisi2,Biroli20}  between the Ising model and the spherical model in the quantum problem too.

~\\
\\
{\bf Acknowledgements}
~\\

We thank Giulio Biroli, Leticia Cugliandolo, Yan Fyodorov, Antoine Georges, Felix Haehl, Nikita Kavokine, Jorge Kurchan, Izabella Lovas, Olivier Parcollet, Marco Schiro, Subhabrata Sen, and Gergely Zar\'and for valuable discussions. This research was supported by the U.S. Department of Energy under Grant DE-SC0019030 and by the Simons Collaboration on Ultra-Quantum Matter which is a grant from the Simons Foundation (651440, S.S.). R.S. is supported by the Princeton Quantum Initiative Fellowship.

\newpage
\appendix

\section{Terms of order $y^1$ in $\mathcal{F}_Q$ for the Ising model}
\label{app:y1}

This appendix contains the $\mc{O}(y)$ corrections to the free energy and saddle-point values from the `quantum' contribution to the free energy, $\mathcal{F}_Q$.

\subsection{Zero longitudinal field}

As in the main text, we first consider the $h=0$ case  of Sec.~\ref{sec:0}.

\subsubsection{Paramagnet}
\label{app:0para}

Here, we continue the analysis of Sec.~\ref{sec:0para}. 

For the saddle point
at order $y^1$, we write
\begin{align}
Q_r (i \nu_n) =  - \frac{\sqrt{\nu_n^2 + \Delta^2 + y \Delta_1^2 - y \Sigma (i \nu)}}{\kappa}\,, \label{Qr1}
\end{align}
where we can choose $\Sigma (0) = 0$ without loss of generality by adjusting the value of $\Delta_1^2$.
Then, from (\ref{rsy3d1}), we obtain
\begin{align}
0 = &-  \left[ y \Delta_1^2 - y \Sigma (i \nu_n) \right] -
 \frac{U}{ \beta} \sum_{\nu_n^\prime} \left[\sqrt{\nu_n^{\prime 2} + \Delta^2 + y \Delta_1^2 - y \Sigma (i \nu_n^\prime)} -  \sqrt{\nu_n^{\prime 2} + \Delta^2 }\right]    \nonumber \\
& + \frac{ 2y }{3\beta^2 \kappa^2}  \sum_{ \nu^{\prime}_n,
\nu^{\prime\prime}_n} \sqrt{\nu_n^{\prime 2} + \Delta^2 }
\sqrt{\nu_n^{\prime\prime 2} + \Delta^2 } \sqrt{(\nu_n -\nu_n^{\prime} - \nu_n^{\prime\prime})^2 + \Delta^2 }  \,.
\label{rsy3d11}
\end{align}
After subtracting from (\ref{rsy3d11}) its value at $\nu_n =0$, we determine 
\begin{align}
\Sigma (i \nu_n) = - \frac{ 2}{3\beta^2 \kappa^2}  \sum_{ \nu^{\prime}_n,
\nu^{\prime\prime}_n}   \sqrt{\nu_n^{\prime 2} + \Delta^2 }
\sqrt{\nu_n^{\prime\prime 2} + \Delta^2 }\left[ \sqrt{(\nu_n -\nu_n^{\prime} - \nu_n^{\prime\prime})^2 + \Delta^2 } - \sqrt{(\nu_n^{\prime} +\nu_n^{\prime\prime})^2 + \Delta^2 } \right]\,. \label{Sigma1}
\end{align} 
Then, $\Delta_1^2$ can be obtained from (\ref{rsy3d11}) as the solution of 
\begin{align}
\Delta_1^2 &= \frac{2}{3\beta^2 \kappa^2}  \sum_{ \nu_n,
\nu^{\prime}_n} \sqrt{\nu_n^{\prime 2} + \Delta^2 }
\sqrt{\nu_n^{2} + \Delta^2 } \sqrt{(\nu_n^{\prime} + \nu_n)^2 + \Delta^2 } + \frac{U}{2 \beta} \sum_{\nu_n}
\frac{\Sigma (i \nu)}{\sqrt{\nu_n^2 + \Delta^2}}
\nonumber \\
&-
 \lim_{y \rightarrow 0} \frac{U}{ y \beta} \sum_{\nu_n} \left[ \sqrt{\nu_n^{2} + \Delta^2 + y \Delta_1^2} -  \sqrt{\nu_n^{2} + \Delta^2 }\right]\,.
\label{rsy3d111}
\end{align}

At the critical point, where $\Delta^2 + y \Delta_1^2 = 0$ and $r=r_{c0} + y r_{c1}$, using (\ref{p2}) and (\ref{p3}), we can write the term proportional to $u$ in (\ref{rsy3d11}) as
\begin{align}
y \Delta_1^2 + y r^{}_{c1} + \frac{U y}{2 \beta} \sum_{\nu_n}
\frac{\left. \Sigma (i \nu_n)\right|_{\Delta=0}}{|\nu_n|}\,,
\end{align}
where $\Sigma (i \nu_n)$ is to be evaluated from (\ref{Sigma1}) at $\Delta =0$. Then, (\ref{rsy3d11}) at $\nu=0$ yields the shift in the position of the critical point at order $y$ as
\begin{align}
r_{c1} &= - \frac{2}{3\beta^2 \kappa^2}  \sum_{ \nu_n,
\nu^{\prime}_n} |\nu_n^{\prime}| |\nu_n| |\nu_n^{\prime} + \nu_n| - \frac{U}{2 \beta} \sum_{\nu_n}
\frac{\left. \Sigma (i \nu_n)\right|_{\Delta=0}}{|\nu_n|}\,.
\label{shiftqc1}
\end{align}

For the free energy at order $y^1$, with $\mathcal{F}_Q = \mathcal{F}_Q^0 + y \mathcal{F}_Q^1$, we can just insert the $\mc{O}(y^0)$ saddle-point values into the terms explicitly dependent upon $y$ in (\ref{FQM}):
\begin{eqnarray}
\mathcal{F}_Q^1 && = - \frac{1}{6 \beta^3 \kappa^4} \sum_{\nu_n, \nu^{\prime}_n,
\nu^{\prime\prime}_n} \sqrt{\nu_n^2 + \Delta^2}\sqrt{\nu_n^{\prime 2} + \Delta^2}\sqrt{\nu_n^{\prime \prime 2} + \Delta^2}\sqrt{(\nu_n+ \nu_n^{\prime}+ \nu_n^{\prime\prime})^2 + \Delta^2}. 
\label{FQMP}
\end{eqnarray}

\subsubsection{Spin glass}
\label{app:0sg}

We continue here the analysis of Sec.~\ref{sec:0sg}.

For the saddle-point equations at order $y^1$, we write
\begin{align}
q_{EA} &= \frac{1}{\kappa U} (r_{c0} - r) + y q_{EA1}, \nonumber \\
Q_{r} (i\nu_n) &= - \frac{|\nu_n|}{\kappa}  + y Q_{r1} ( i \nu_n). \label{qea11}
\end{align}
Then, Eqs.~(\ref{ifinaleq1}) and (\ref{ifinaleq2}) yield
\begin{alignat}{2}
0 &=   &&2 |\nu_n| Q_{r1} (i \nu_n)  +
 \frac{U}{\beta} \sum_{\nu_n^\prime} Q_{r1} ( i\nu_n^\prime ) + u q_{EA1} + \frac{ 2}{3 \kappa^3 \beta^2}  \sum_{ \nu^{\prime}_n,
\nu^{\prime\prime}_n} |\nu^{\prime}_n| |\nu^{\prime\prime}_n| |\nu_n -
\nu^{\prime}_n-\nu^{\prime\prime}_n| \nonumber \\
&  &&- \frac{2}{\kappa^2\beta} q_{EA0} \sum_{\nu_n^\prime} |\nu_n^\prime| |\nu_n - \nu_n^\prime| - 
2 Q_{r 1} (0) |\nu_n|, \label{finaleq1a} \\
q_{EA0}^2 & = &&- \kappa Q_{r1} (0). \label{finaleq2ay} 
\end{alignat}
From (\ref{finaleq1a},\ref{ifinaleq2}), we determine that 
\begin{align}
q_{EA1}  = -\frac{1}{\beta} \sum_{\nu_n} Q_{r1} ( i\nu_n ) - \frac{ 2}{3 \kappa^3 U \beta^2}  \sum_{ \nu_n,
\nu^{\prime}_n} |\nu^{\prime}_n| |\nu_n| |
\nu^{\prime}_n+\nu_n| + \frac{2}{\kappa^2 U \beta} q_{EA0} \sum_{\nu_n} |\nu_n|^2,
\label{finaleq5} 
\end{align}
and
\begin{align}
 Q_{r1} (i \nu_n) & =   - \frac{q_{EA0}^2}{\kappa} - \frac{ 1}{3 \kappa^3 \beta^2 |\nu_n|}  \sum_{ \nu^{\prime}_n,
\nu^{\prime\prime}_n} \Bigl( |\nu^{\prime}_n| |\nu^{\prime\prime}_n| |\nu_n -
\nu^{\prime}_n-\nu^{\prime\prime}_n| -  |\nu^{\prime}_n| |\nu^{\prime\prime}_n| |
\nu^{\prime}_n+\nu^{\prime\prime}_n| \Bigr) \nonumber \\
&~~~~ + \frac{1}{\beta \kappa^2 |\nu_n|} q_{EA0} \sum_{\nu_n^\prime} \Bigl( |\nu_n^\prime| |\nu_n - \nu_n^\prime| - |\nu_n^\prime|^2 \Bigr) \,. \label{finaleq3a} 
\end{align}
At $T=0$, we obtain, with a frequency cutoff $|\nu_n| < \Lambda$,
\begin{align}
 Q_{r1} (i \nu_n) & =   - \frac{q_{EA0}^2}{\kappa} 
 - \frac{1}{6 \pi \kappa^3} \Bigl( \frac{2 \Lambda^3}{3} |\nu_n| - \frac{\Lambda^2}{3} |\nu_n|^2  + \frac{1}{20} |\nu_n|^4 \Bigr) 
+ \frac{1}{6\pi \kappa^2 } q_{EA0} |\nu_n|^2.
 \label{finaleq3aa} 
\end{align}

The shift in the quantum critical point is obtained by setting $r = r_{c0} + y r_{c1}$ in (\ref{finaleq5}) and (\ref{finaleq3a}), and this yields the same value of $r_{c1}$ as that obtained from the vanishing gap condition on the paramagnetic side in (\ref{shiftqc1}).

For the free energy 
at order $y^1$, we now obtain from (\ref{FQM})
\begin{equation}
\mathcal{F}_Q^1  = - \frac{r q^{}_{EA1}}{\kappa} -  \frac{ q_{EA0}^2}{\beta \kappa^2} \sum_{\nu_n} \nu_n^2
+ \frac{2 q^{}_{EA0}}{3 \kappa^3 \beta^2} \sum_{\nu_n, \nu^{\prime}_n} |\nu_n| |\nu_n^{\prime}||\nu_n+ \nu_n^{\prime}|
- \frac{ 1}{6 \beta^3 \kappa^4} \sum_{\nu_n, \nu^{\prime}_n,
\nu^{\prime\prime}_n} |\nu_n| |\nu_n^{\prime}||\nu_n^{\prime\prime}||\nu_n+ \nu_n^{\prime}+ \nu_n^{\prime\prime}|. 
\label{FQMRSB}
\end{equation}

\subsection{Nonzero longitudinal field}

Next, we consider the case of $h\neq 0$ studied in Sec.~\ref{sec:h}.

\subsubsection{Replica-symmetric solution}
\label{app:hrs}

This section continues the analysis of Sec.~\ref{sec:hrs}.

For the saddle point at order $y^1$, we write as in (\ref{Qr1})
\begin{align}
Q_r (i \nu_n) &=  - \frac{\sqrt{\nu_n^2 + \Delta^2 + y \Delta_1^2 - y \Sigma (i \nu)}}{\kappa}, \nonumber \\
q_{EA} & = q_{EA0} + q_{EA1}, \label{Qr1a}
\end{align}
where we can again choose $\Sigma (0) = 0$ without loss of generality by adjusting the value of $\Delta_1^2$.
Then, from (\ref{rsy3dd}), we obtain
\begin{align}
0 = &-  \left[ y \Delta_1^2 - y \Sigma (i \nu_n) \right] -
 \frac{U}{ \beta} \sum_{\nu_n^\prime} \left[\sqrt{\nu_n^{\prime 2} + \Delta^2 + y \Delta_1^2 - y \Sigma (i \nu_n^\prime)} -  \sqrt{\nu_n^{\prime 2} + \Delta^2 }\right]    \nonumber \\
&+ U \kappa y q^{}_{EA1} + \frac{ 2y }{3\beta^2 \kappa^2}  \sum_{ \nu^{\prime}_n,
\nu^{\prime\prime}_n} \sqrt{\nu_n^{\prime 2} + \Delta^2 }
\sqrt{\nu_n^{\prime\prime 2} + \Delta^2 } \sqrt{(\nu_n -\nu_n^{\prime} - \nu_n^{\prime\prime})^2 + \Delta^2 } \nonumber \\
&- \frac{2 y}{\beta \kappa} q^{}_{EA0} \sum_{\nu_n^\prime} \sqrt{\nu_n^{\prime 2} + \Delta^2 }
 \sqrt{(\nu_n -\nu_n^{\prime} )^2 + \Delta^2} +
2 y q_{EA0}^2 \sqrt{\nu_n^{2} + \Delta^2 }  \,.
\label{rsy3dd1}
\end{align}
After subtracting from (\ref{rsy3dd1}) its value at $\nu_n =0$, we determine 
\begin{align}
 \Sigma (i \nu_n)  = &- \frac{ 2}{3\beta^2 \kappa^2}  \sum_{ \nu^{\prime}_n,
\nu^{\prime\prime}_n}   \sqrt{\nu_n^{\prime 2} + \Delta^2 }
\sqrt{\nu_n^{\prime\prime 2} + \Delta^2 }\left[ \sqrt{(\nu_n -\nu_n^{\prime} - \nu_n^{\prime\prime})^2 + \Delta^2 } - \sqrt{(\nu_n^{\prime} +\nu_n^{\prime\prime})^2 + \Delta^2 } \right]  \nonumber \\
&+ \frac{2 }{\beta \kappa} q_{EA0} \sum_{\nu_n^\prime} \left[ \sqrt{\nu_n^{\prime 2} + \Delta^2 }
 \sqrt{(\nu_n -\nu_n^{\prime} )^2 + \Delta^2} - (\nu_n^{\prime 2} + \Delta^2) \right] -
2 q_{EA0}^2 \left[\sqrt{\nu_n^{2} + \Delta^2 } - \Delta \right] \,. \label{Sigma1a}
\end{align} 
Then, $\Delta_1^2$ obeys, from (\ref{rsy3dd1}),  
\begin{align}
\Delta_1^2 &= U \kappa q^{}_{EA1} + \frac{ 2y }{3\beta^2 \kappa^2}  \sum_{ \nu^{\prime}_n,
\nu^{\prime\prime}_n} \sqrt{\nu_n^{\prime 2} + \Delta^2 }
\sqrt{\nu_n^{\prime\prime 2} + \Delta^2 } \sqrt{(\nu_n -\nu_n^{\prime} - \nu_n^{\prime\prime})^2 + \Delta^2 } \nonumber \\
&- \frac{2 y}{\beta \kappa} q^{}_{EA0} \sum_{\nu_n^\prime} \sqrt{\nu_n^{\prime 2} + \Delta^2 }
 \sqrt{(\nu_n -\nu_n^{\prime} )^2 + \Delta^2} +
2 y q_{EA0}^2 \sqrt{\nu_n^{2} + \Delta^2 } 
\nonumber \\
& + \frac{U}{2 \beta} \sum_{\nu_n}
\frac{\Sigma (i \nu)}{\sqrt{\nu_n^2 + \Delta^2}} -
 \lim_{y \rightarrow 0} \frac{U}{ y \beta} \sum_{\nu_n} \left[ \sqrt{\nu_n^{2} + \Delta^2 + y \Delta_1^2} -  \sqrt{\nu_n^{2} + \Delta^2 }\right]\,.
\label{rsy3dd11}
\end{align}
Determination of $q_{EA1}$ and $\Delta_1$ requires solution of (\ref{rsy3dd1}), along with a second equation obtained from (\ref{rs1})
\begin{align}
 -3 q^{}_{EA0} \frac{\Delta_1^2}{2 \Delta} - 3 \Delta q^{}_{EA1}  + q_{EA0}^3  = 0 \,.\label{rs1z}
\end{align}

For the free energy at order $y^1$, we now obtain from (\ref{FQM})
\begin{alignat}{1}
    \mathcal{F}_Q^1  = &-  \frac{y q_{EA0}^2}{\beta \kappa^2} \sum_{\nu_n} (\nu_n^2 + \Delta^2)
+ \frac{2 y q^{}_{EA0}}{3 \kappa^3 \beta^2} \sum_{\nu_n, \nu^{\prime}_n} \sqrt{\nu_n^2 + \Delta^2}\sqrt{\nu_n^{\prime 2} + \Delta^2}\sqrt{(\nu_n+ \nu_n^{\prime})^2 + \Delta^2}
\nonumber
\\ & - \frac{ y}{6 \beta^3 \kappa^4} \sum_{\nu_n, \nu^{\prime}_n,
\nu^{\prime\prime}_n} \sqrt{\nu_n^2 + \Delta^2}\sqrt{\nu_n^{\prime 2} + \Delta^2}\sqrt{\nu_n^{\prime \prime 2} + \Delta^2}\sqrt{(\nu_n+ \nu_n^{\prime}+ \nu_n^{\prime\prime})^2 + \Delta^2}. 
\label{FQMRS}
\end{alignat}

\section{Classical spherical $p$-rotor model}

In the classical limit $g=0$, all components of $Q_{ab} (\tau)$ and $\Sigma_{ab} (\tau)$ are independent of $\tau$, and the theory should reduce to that in Ref.~\onlinecite{Crisanti_statics}.
So, we have $Q_{ab} (\omega_n) = \beta Q_{ab} (\tau) \delta_{\omega_n, 0} \equiv \beta Q_{ab} \delta_{\omega_n, 0}$, and similarly for $\Sigma_{ab}$ and $\lambda_{ab}$. Then, the effective action in (\ref{eq:Seff2}) becomes
\begin{alignat}{1}
\label{eq:Seff2a}
	\frac{S_{\rm eff}}{N}&=\frac{1}{2} \log\det \left[\frac{1}{\pi} \left( -i\delta_{ab}  z^a  +   i \beta \lambda^{}_{ab} \right)\right]+i \beta \sum_{a=1}^n z^a
 \nonumber\\
 &-\sum_{a,b=1}^n \Bigg( \beta^2 \,i\lambda_{ab} Q_{ab} + \frac{\beta h^2}{4} \bigg[ -i \delta_{ab}  z^a  +   i \beta \lambda^{}_{ab} \bigg]^{-1}_{ab}  +\frac{\beta^2 J^2 }{4}  \left(Q^{}_{ab}\right)^p \Bigg)~.
\end{alignat}
Let us now change our notation a bit to obtain expressions similar to those in Ref.~\onlinecite{QPM}. We define
\begin{align}
- iz^a  = {\overline{z}} \quad , \quad i \lambda_{ab}  = -{\sigma_{ab} }.
\end{align}
Then, the effective action is
\begin{alignat}{1}
\label{eq:Seff3a}
	&\frac{S_{\rm eff}}{N}=\frac{1}{2} \log\det \left[\frac{1}{\pi } \left( \delta_{ab}  \overline{z}  -\beta \sigma^{}_{ab} \right)\right] - \beta \sum_{a=1}^n {\overline{z}}
  + \sum_{a,b=1}^n \Bigg( \beta^2 \,\sigma_{ab} Q_{ab} - \frac{\beta h^2}{4} \bigg[ \overline{z} \delta_{cd}   - \beta \sigma^{}_{cd} \bigg]^{-1}_{ab}  -\frac{\beta^2 J^2 }{4}  \left(Q^{}_{ab}\right)^p \Bigg) ~.
\end{alignat}
Now, as before, we use the identity (\ref{CSidentity}) to absorb the $h^2$ term into the $\log \det$
\begin{alignat}{1}
\label{eq:Seff3b}
	&\frac{S_{\rm eff}}{N}=\frac{1}{2} \log\det \left[\frac{1}{\pi } \left( \delta_{ab}  \overline{z}  -\beta \sigma^{}_{ab} - \frac{\beta h^2}{2}  \right)\right] - \beta \sum_{a=1}^n {\overline{z}}
 + \sum_{a,b=1}^n \Bigg( \beta^2 \,\sigma_{ab} Q_{ab} -\frac{\beta^2 J^2 }{4}  \left(Q^{}_{ab}\right)^p \Bigg) ~.
\end{alignat}
The saddle-point equation with respect to $\sigma_{ab}$ is 
\begin{align}
\beta Q_{ab} & = \frac{1}{2} \bigg[ \delta_{ab}  \overline{z} -\beta \sigma^{}_{ab} - \frac{\beta h^2}{2}\bigg]^{-1}_{ab}  .\label{finaleq2} 
\end{align}
Inserting this back into (\ref{eq:Seff3b}), we obtain an effective action just for $Q_{ab}$ and $\overline{z}$,
\begin{alignat}{1}
\label{eq:Seff3c}
	&\frac{S_{\rm eff}}{N}= - \frac{1}{2} \log\det Q - \beta \sum_{a=1}^n {\overline{z}}
  + \sum_{a,b=1}^n \Bigg( \beta \overline{z} \delta_{ab} Q_{ab} - \frac{\beta^2 h^2}{2} Q_{ab}  -\frac{\beta^2 J^2 }{4}  \left(Q^{}_{ab}\right)^p \Bigg) ~.
\end{alignat}
Then, the remaining saddle-point equations are
\begin{align}
1 & = Q_{aa}, \nonumber \\
 Q^{-1}_{ab} & =  2 \beta \overline{z} \delta^{}_{ab} - \beta^2 h^2   -\frac{p\beta^2 J^2 }{2}  \left(Q^{}_{ab}\right)^{p-1}. \label{finaleq3} 
\end{align}

\subsection{Replica-symmetric solution}

Now, we use the replica-symmetric ansatz 
\begin{align}
    Q_{ab} &=  q  + Q_r \delta_{ab}, \nonumber \\
      \sigma_{ab}  &=  \varrho  + \sigma_r \delta_{ab}. 
    \label{Qansatzc}
\end{align}
Then, using the identities in Appendix~\ref{app:replica}, the equations in (\ref{finaleq3}) become
\begin{align}
1 & = q +  Q_r, \label{e1ap} \\
\sigma_r  & =  \frac{pJ^2}{4} (q + Q_r)^{p-1} -  \frac{pJ^2}{4} q^{p-1}, \label{e2ap} \\
\varrho  & =  \frac{pJ^2}{4} q^{p-1}, \label{e3ap} \\
\frac{1}{\beta Q_r} & = 2 (\beta{\overline{z}} - \beta \sigma_r ), \label{e4ap} \\
 \frac{q}{Q_r^2} &=    \beta^2 h^2   + \frac{p\beta^2 J^2 }{2}  q^{p-1}.  
  \label{e5ap} 
\end{align}

\begin{figure}[tb]
\center{

\includegraphics[width=3.15in]{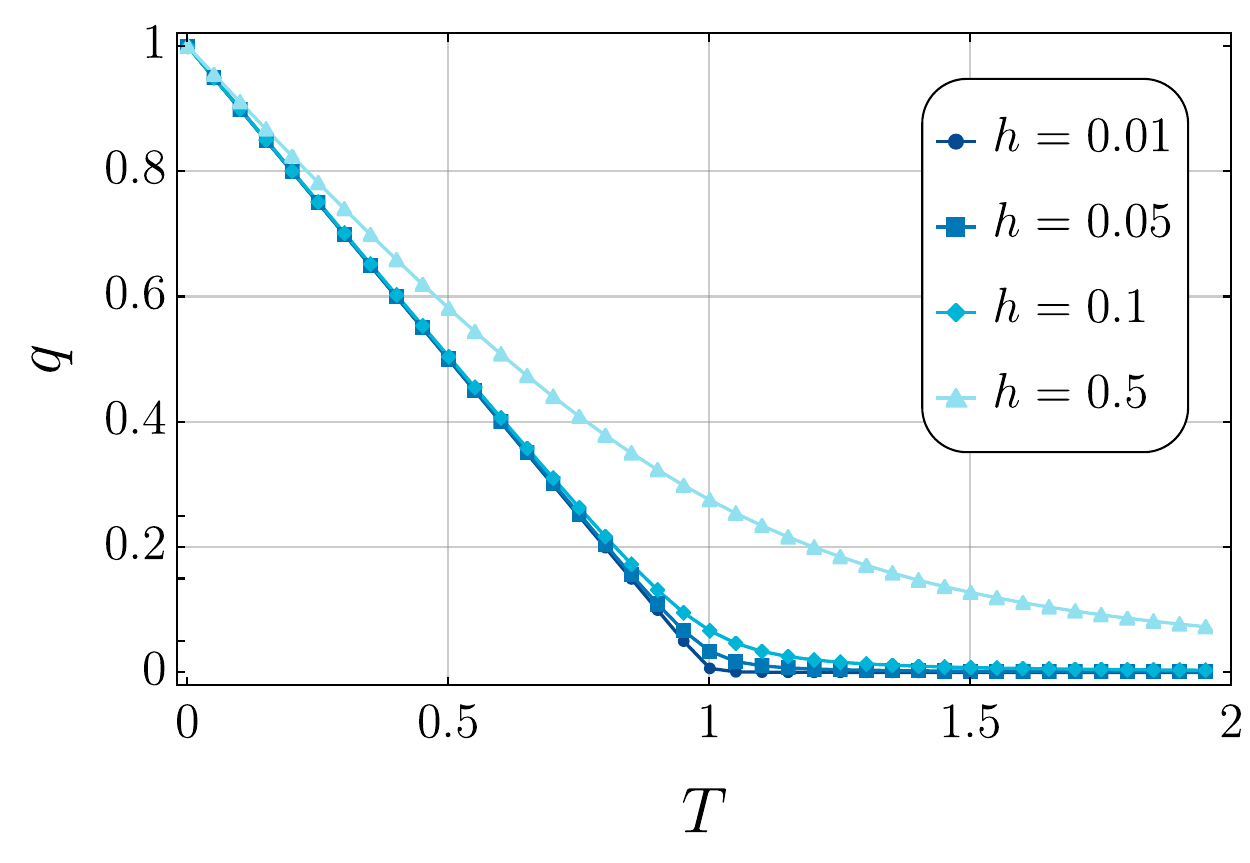}}
\caption{Spin glass order parameter $q$ of the classical spherical $p=2$-rotor model (with $J$\,$=$\,$1$), as a function of the temperature $T$ for several values of the longitudinal field $h$. For the replica-symmetric solution considered here, the spin glass phase transition occurs at $T=1$.}
\label{fig:clas_p}
\end{figure}

Focusing on $p=2$ first, we can simplify the above to
\begin{align}
1 & = q +  Q_r,  \label{e1aaa} \\
    \beta Q_r & =  \frac{1}{\overline{z} + \sqrt{\overline{z}^2 - J^2}}, \label{e4aaa} \\
      q & =  \frac{J^2 q + h^2}{(\overline{z} + \sqrt{\overline{z}^2 - J^2})^2}. \label{e5aaa} 
\end{align}
These equations are easily seen to be equivalent to the classical limit of Eqs. (\ref{qq0},\ref{qq1},\ref{qq2}) derived for the quantum model later. Figure~\ref{fig:clas_p} showcases the temperature dependence of the order parameter $q$ for nonzero $h$. For $p$\,$=$\,$2$, $h$\,$=$\,$0$, it is known \cite{kosterlitz1976spherical} that the replica-symmetric solution is stable and optimal, and we expect that this statement continues to hold for any $h\neq 0$ as well \cite{Crisanti_statics}. 

\subsection{One-step replica symmetry breaking}

The classical spherical $p$-rotor model can be solved exactly for all $p$, temperatures, and fields, and hosts a one-step RSB phase for any $p>2$ \cite{Crisanti_statics} (akin to the model without the spherical constraint \cite{Gardner85}). In this case, for the diagonal elements of $Q_{ab}$ and $\sigma_{ab}$, we use the ansatz
\begin{align}
    Q_{aa} &=  q_1  + Q_r,   \nonumber \\
    \sigma_{aa} & = \varrho_1 + \sigma_r, \label{Qansatz2c}
\end{align}
while the off-diagonal elements are as in (\ref{q1_rho1_rsb}).
The equations analogous to (\ref{e1a})--(\ref{e5a}) are
\begin{align}
1 & = q_1 +  Q_r,  \label{cl:e1bb} \\
\sigma_r  & =  \frac{pJ^2}{4} (q_1 + Q_r)^{p-1} -\frac{pJ^2}{4} q_1^{p-1}, \label{e2bb} \\
\varrho_1 & =  \frac{pJ^2}{4} q_1^{p-1}, \label{cl:e3bb} \\
 \varrho_0 & =  \frac{pJ^2}{4} q_0^{p-1}, \label{cl:e4bb} \\
 \frac{1}{\beta Q_r} & = 2 (\overline{z} - \beta \sigma_r ), \label{cl:e5bb} \\
 \frac{q_1 Q_r + x(q_1-q_0)^2}{Q_r (Q_r + x (q_1 - q_0))^2} &=    \beta^2 h^2   + \frac{p\beta^2 J^2 }{2}  q_1^{p-1}, \label{cl:e6bb} \\ 
 \frac{q_0}{(Q_r + x (q_1 - q_0))^2} &=    \beta^2 h^2   + \frac{p\beta^2 J^2 }{2}  q_0^{p-1}.  
 \label{cl:e7bb}
\end{align}

We are mainly interested in the $\beta \rightarrow \infty$ limit. From the equations above,  we see that this limit exists if we write
\begin{align}
    q_1  = \widehat{q} - \frac{\theta}{\beta}, \quad
    x  = \frac{\mu}{\beta}  \,,
\end{align}
where $\widehat{q}$ is a $\beta$-independent constant, and $\theta$ and $\mu$ are finite as $\beta \rightarrow \infty$. For the present classical theory, $\widehat{q} = 1$, but we will see that $\widehat{q} < 1$ in the $g \neq 0$ theory.

\subsubsection{Gapless condition}

The gapless condition, obtained in the $g \rightarrow 0$ limit of (\ref{gps4}) is
\be
\frac{1}{\beta^2 Q_r^2} = \frac{p(p-1) J^2}{2} q_1^{p-2}\,. \label{gps4c}
\ee

\subsubsection{Free energy}

The classical limit of the action (\ref{eq:Seff40}) is
\begin{alignat}{1}
\label{eq:Seff50}
	\frac{S_{\rm eff}}{Nn}= &-\frac{1}{2} \log[Q_r] 
 - \frac{1}{2} \frac{ q_0}{Q_r + x (q_1 - q_0)} + \frac{1}{2x} \ln \frac{ Q_r}{Q_r + x (q_1 - q_0)}  \\
 & \nonumber
 -  \beta \overline{z}
  +  \beta \overline{z} (Q_r + q_1)
  - \frac{\beta^2 h^2}{2}\left[Q_r +  x(q_1 - q_0) 
 \right] -\frac{\beta^2 J^2 }{4}   \left[ \left(Q_r + q_1\right)^p - q_1^p (1-x) - q_0^p x \right]. 
 \end{alignat}
We can now confirm that the equations (\ref{cl:e1bb})--(\ref{cl:e7bb}) are indeed the saddle-point equations obtained from Eq.~(\ref{eq:Seff50}).

 The stationarity condition with respect to the breakpoint yields
 \begin{align}
    \frac{\partial S_{\rm eff}}{\partial x} = 0 
     \quad \Rightarrow \quad &
     \frac{J^2 \beta^2}{4} (q_1^p - q_0^p) - \frac{1}{2x^2}\log \left[ 1 + \frac{x (q_1-q_0)}{ Q_r}\right] \nonumber \\ & + \frac{\beta^2 h^2}{2} (q_1 - q_0) + \frac{ (q_1 - q_0)(Q_r + q_1 x - 2 q_0 x)}{2x (Q_r + x (q_1 - q_0))^2} = 0.
     \label{dSdx2}
 \end{align}
One can use either (\ref{gps4c}) or (\ref{dSdx2}) to determine $x$.

\section{Replica-symmetric solution for the $h\ne 0$ quantum spherical $p=2$ model}
\label{sec:p2spherical}

We now consider the quantum version of the spherical $p$\,$=$\,$2$-rotor model in a longitudinal field $h$. As discussed in Ref.~\onlinecite{YSR93}, for $h=0$, the replica-symmetric solution is the optimal one. Anticipating the same conclusion to hold for the case of $h\neq0$, in what follows, we focus on the replica-symmetric solution. 

For $p$\,$=$\,$2$, the equations \eqref{finaleq}, combined with the replica-symmetric ansatz \eqref{Qansatz}, reduce to
\begin{align}
1 & = q + \frac{1}{\beta} \sum_{\omega_n} Q_r (\omega_n),  \label{qq0} \\
\varrho  & =  {gJ^2} q,  \\
    Q_r (\omega_n) & = \frac{g}{\omega_n^2 + \overline{\lambda} - gJ^2Q_r(\omega_n)},  \\
      q & = \frac{g \varrho +  g^2 h^2}{[\overline{\lambda} - gJ^2Q_r (\omega_n=0)]^2}. 
\end{align}
Specifically, the equation for $Q_r(\omega_n)$ can be written as
\begin{align}
 Q_r (\omega_n)  &= \frac{2g}{\omega_n^2+\overline{\lambda}+\sqrt{(\omega_n^2 +\overline{\lambda})^2-4g^2J^2}}, \label{qq1} \\
 Q_r(\omega_n=0) &= \frac1{2gJ^2}\left(\olambda - \sqrt{\olambda^2-4g^2J^2}\right).
\end{align}
Then, the order parameter is given by 
\begin{align}
q=4g^2\frac{J^2q+h^2}{\left[\olambda+\sqrt{\olambda^2-4g^2J^2}\right]^2}. \label{qq2}
\end{align}
From the equation above, we observe  that the condition $\olambda \ge 2gJ$ has to be satisfied for the order parameter to be a real number. 
Note that for $h=0$, we find
\begin{align}
\olambda|_{h=0} = 2Jg,
\end{align}
which is consistent with Eq.~(33.41) of Ref.~\onlinecite{QPM}.
For $h\neq0$, the equation for the order parameter can be formulated as
\begin{align}
q=\frac{4g^2h^2}{\left[\olambda+\sqrt{\olambda^2-4g^2J^2}\right]^2 - 4g^2 J^2} = \frac{2g^2h^2}{\olambda^2 - 4g^2J^2+\olambda\sqrt{\olambda^2-4g^2J^2}}.
\end{align}
We can simplify the expression above further to
\begin{align}\label{eq:q}
q = -\frac{h^2}{2J^2}\frac{\olambda^2-4g^2J^2-\olambda\sqrt{\olambda^2-4g^2J^2}}{\olambda^2-4g^2J^2} = -\frac{h^2}{2J^2}\frac{\lambda^{*2}-\olambda\lambda^*}{\lambda^{*2}} = -\frac{h^2}{2J^2}\frac{\lambda^*-\olambda}{\lambda^*} = \frac{h^2}{2J^2}\left(\frac{\olambda}{\lambda^*}-1\right),
\end{align}
where we introduce the notation
\begin{align}
\lambda^{*2} = \olambda^2-4g^2J^2.
\end{align}
We note that the order parameter has to be positive, $q>0$, so $\olambda>\lambda^*$. In addition, we require $\olambda>2gJ$ such that $\lambda^*>0$.

\begin{figure}
\center{
(a)\includegraphics[width=3.15in]{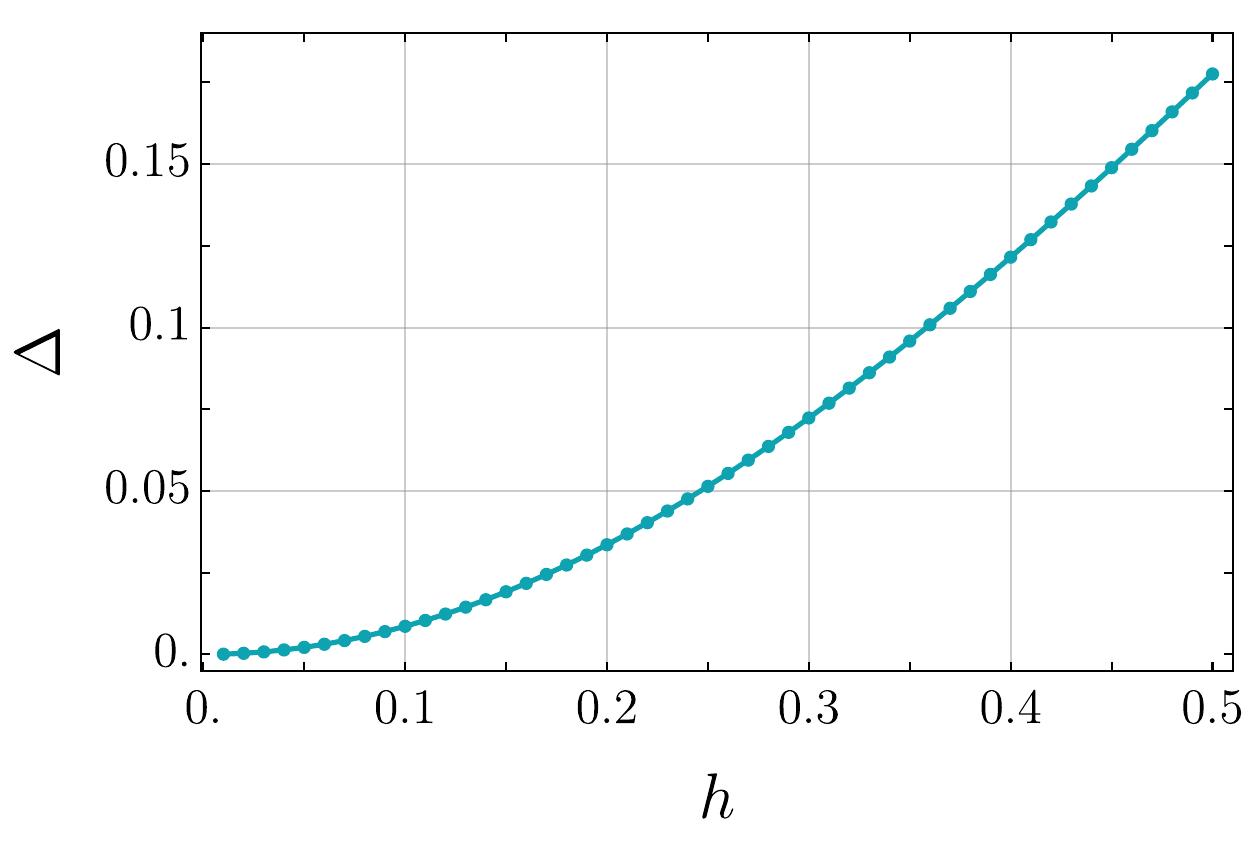}\,\,\,\,\,\,
(b)\includegraphics[width=3.15in]{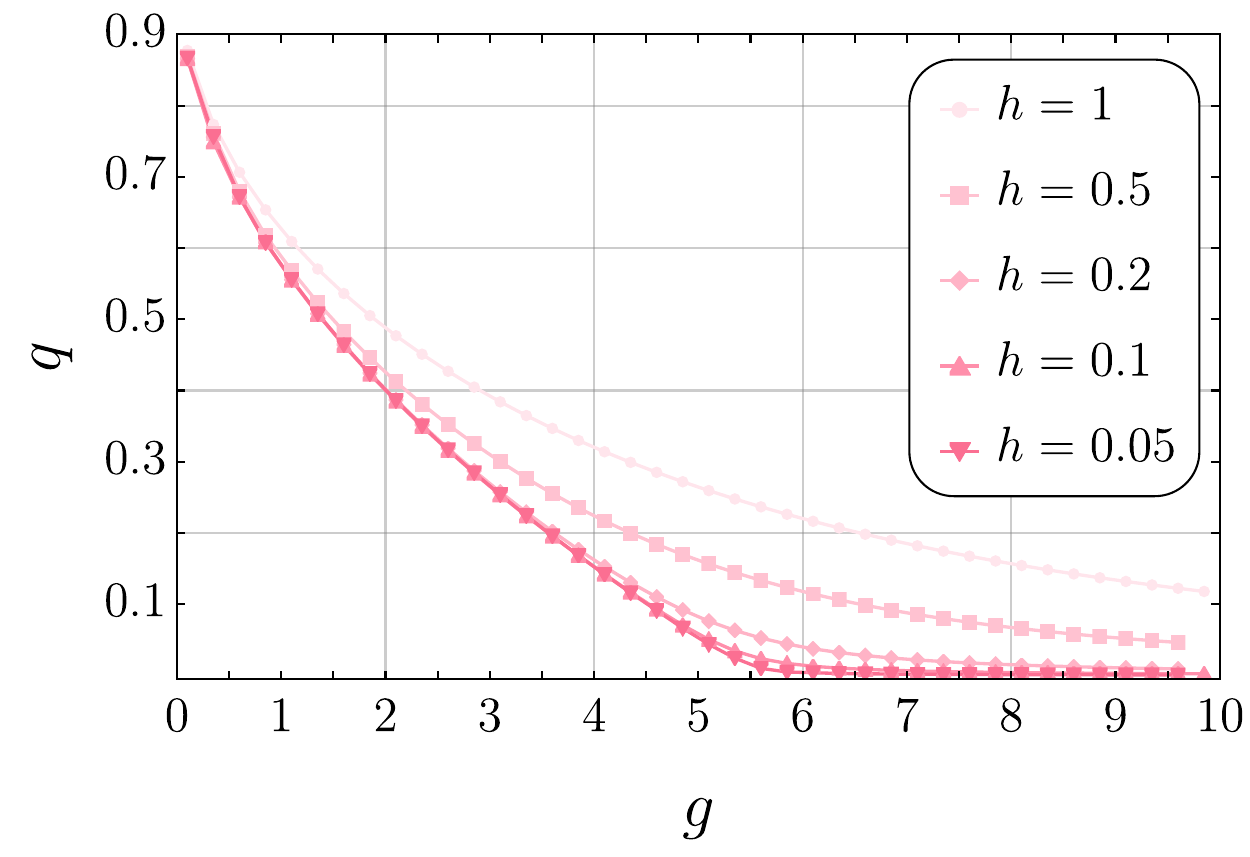}}
\caption{(a) Gap in the spectrum \eqref{eq:gapp2} as a function of the longitudinal field strength $h$ for the quantum spherical  $p=2$-rotor model, taking $g$\,$=$\,$1$. The system is gapless in the limit $h\to 0$. (b) The order parameter $q$ as a function of the transverse magnetic field for several values of $h$, illustrating the absence of a sharp phase transition at any nonzero $g$ and $h$. 
}
\label{fig:deltah_qg}
\end{figure}

Finally, we rewrite the correlator $Q_r(\omega_n)$ in terms of our new notation as
\begin{align}
Q_r (\omega_n)  &= \frac{2g}{\omega_n^2+\overline{\lambda}+\sqrt{\omega_n^4 + 2\omega_n^2\olambda+{\lambda}^{*2}}}.
\end{align}
Since $\lambda^*>0$, there is always a gap in the spectrum. The gap is given by the value of the frequency at which the square root in the Green's function (analytically continued to the real frequency) changes sign. This value is simply 
\begin{align}
\Delta &=\sqrt{\olambda - 2gJ}, 
\label{eq:gapp2}
\end{align}
which is always greater than zero in the presence of a nonzero field $h$. The resulting behavior of the gap as a function of the longitudinal field is shown in Fig.~\ref{fig:deltah_qg}(a).

To find the value of the spin glass order parameter, we need to find $\olambda$ according to
\begin{align}
q&= \frac{h^2}{2J^2}\left(\frac{\olambda}{\lambda^*}-1\right),\\
Q_r (\omega_n)  &= \frac{2g}{\omega_n^2+\overline{\lambda}+\sqrt{\omega_n^4 + 2\omega_n^2\olambda+{\lambda}^{*2}}},\\
1 & = q + \frac{1}{\beta} \sum_{\omega_n} Q_r (\omega_n).
\end{align}
The numerical solution for $q$ as a function of $g$ is shown in Fig.~\ref{fig:deltah_qg}(b). For $h=0$, one can also compute the critical point, $g_c$, analytically to find 
\begin{align}
g_c = \frac{9\pi^2 J}{16}\approx 5.55 J,
\end{align}
which reproduces the result obtained in Ref.~\onlinecite{QPM}.

\section{Replica-symmetric solution for the $h = 0$ quantum spherical $p=3$ model at large $g$}
\label{sec:largeg}
\begin{figure}[b]
(a)\includegraphics[width=3.07in]
{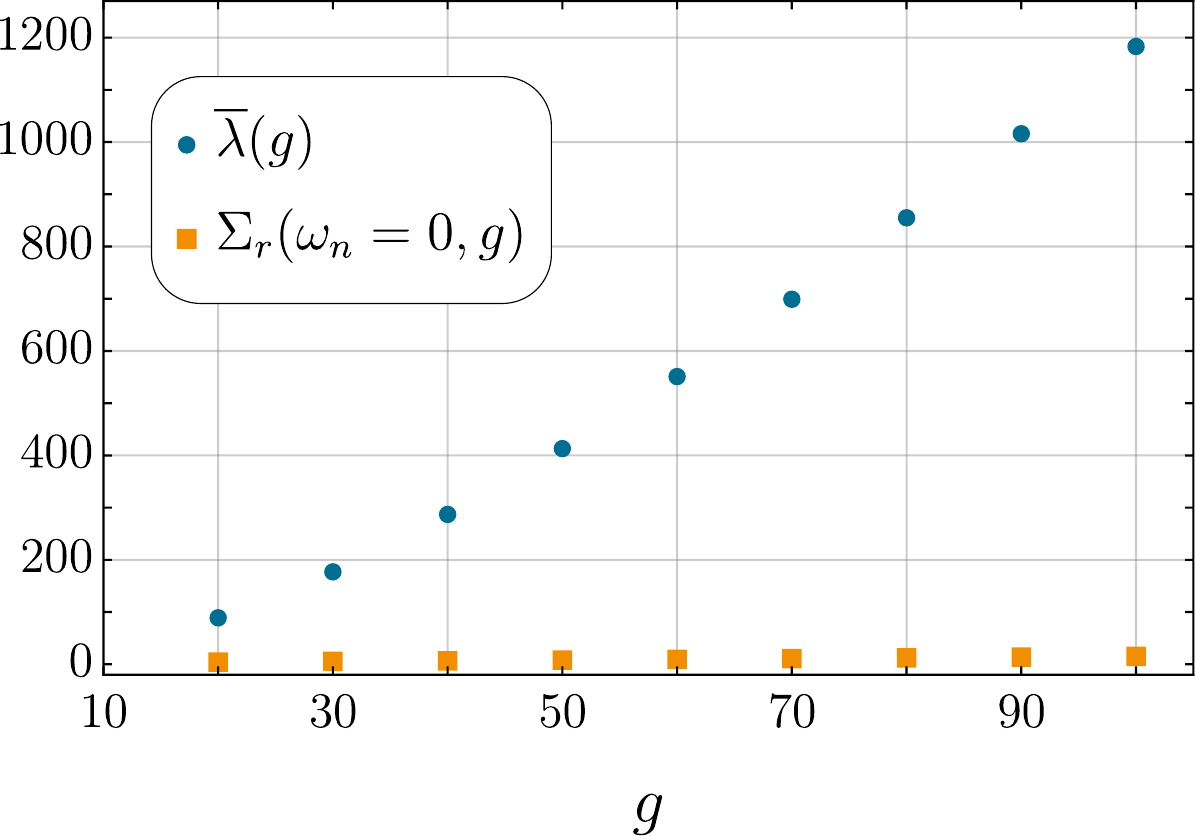}\,\,\,\,\,\,
(b)\includegraphics[width=3.15in]{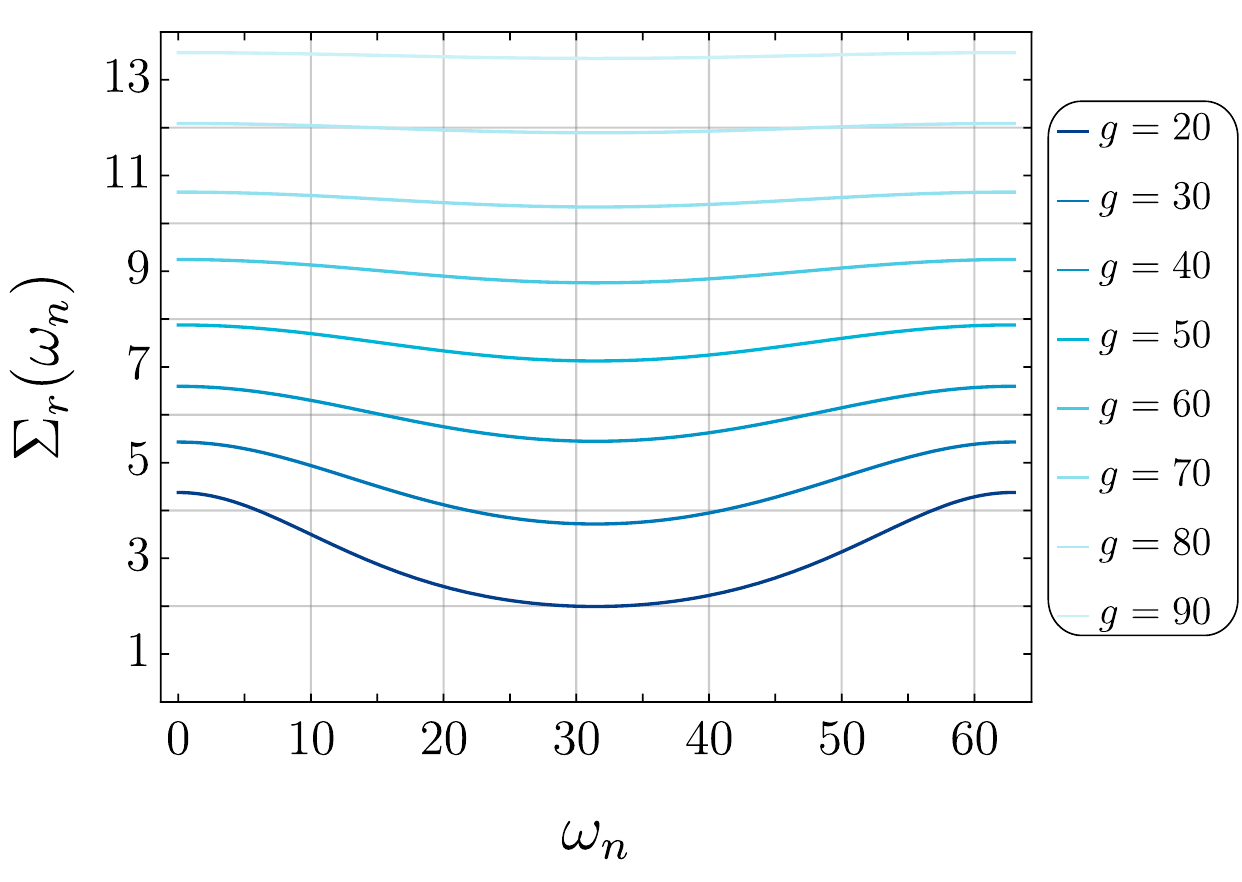}
\caption{(a) Behavior of $\olambda$ and $\Sigma_r(\omega_n=0)$ as a function of $g$ for the quantum spherical $p=3$-rotor model; $\olambda$ grows significantly faster with $g$ than $\Sigma_r$. In addition, $\Sigma_r(\omega_n=0)\ll\overline{\lambda}$, which justifies the leading-order approximation made in \eqref{eq:para_approx}. (b) Self-consistent solution of the paramagnetic equations \eqref{eq:para1}--\eqref{eq:para2} for the self energy, $\Sigma_r$, for large values of $g$. The self energy becomes a constant in the limit $g\to \infty$. }
\label{fig:para_Sigma}
\end{figure}

 In the limit of large longitudinal fields $g$, one can analytically solve the equations \eqref{e1a}--\eqref{e5a}  at $h=0$. We first note that in this limit, the order parameter $q$ is zero, {\it i.e.\/}, the system is in a paramagnetic phase as shown in Fig.~\ref{fig:quan_p3_RS}. To obtain the solution in this phase, we rewrite the saddle-point  equations in the following form: 
\begin{align}
1 & = Q_r(\tau=0), \label{eq:para1} \\
\Sigma_r (\tau) & =   \frac32 gJ^2Q_r^2(\tau),\\
    Q_r (\omega_n) & = \frac{g}{\omega_n^2 + \overline{\lambda} - \Sigma_r (\omega_n)}. \label{eq:para2}
\end{align}
We numerically solve these equations for several values of $g$ and infer that $\lambda$ grows faster with $g$ than $\Sigma_r$, which can be seen  in Fig.~\ref{fig:para_Sigma}(a). In addition, $\Sigma_r(\omega_n)$ is close to a constant in this limit as evidenced by Fig.~\ref{fig:para_Sigma}(b). We can therefore assume that, to leading order,
\begin{align}
    Q_r (\omega_n) & = \frac{g}{\omega_n^2 + \overline{\lambda} }. \label{eq:para_approx} 
\end{align}
The resulting analytical solution yields 
\begin{align}
\olambda=\frac{g^2}{4},\,\,\,Q_r(\omega_n)=\frac{g}{\omega_n^2+\olambda},\,\,\,\Sigma_r(\tau)=\frac{3g}{2}e^{-g|\tau|},\,\,\,\Sigma_r(\omega_n)=\frac{3 g^2 }{g^2+\omega_n^2}\approx \frac{3 }{g^2}(g^2-{\omega_n^2}).
\end{align}

To improve upon our earlier approximation, we compute the Green's function to second order in $1/g$. At this order, we obtain the following result: 
\begin{align}
\olambda&=\frac{g^2}{4}+\frac94,\\
Q_r(\omega_n)&=\frac{g}{\omega_n^2(1+3/g^2)+(\olambda-3)},\\
\Sigma_r(\tau)&=\frac{3g^5}{2(g^4-9)}e^{-g|\tau|\sqrt{\frac{g^2-3}{g^2+3}}},\\
\Sigma_r(\omega_n)&=\frac{3 g^6 }{\sqrt{g^4-9}}\frac1{g^2(g^2-3)+\omega_n^2(g^2+3)}.
\end{align}
We therefore conclude that at large $g$, the spectral function has the form of a delta function
\begin{align}
\rho(\omega)=\frac{g^3}{g^2+3}\delta\left(\omega^2-\frac{g^2}4\frac{g^2-3}{g^2+3}\right),
\end{align}
with a finite gap. Thus, within this approximate framework, we expect the replica-symmetric solution to persist up to $g\approx \sqrt{3}$.

\section{Replica identities}
\label{app:replica}

We consider $n \times n$ replica matrices $A_{ab}$ whose off-diagonal elements are parametrized by the Parisi function $a(u)$, $u \in [0,1]$, and the diagonal element $A_{aa} = \tilde{a}$ (similarly for $B_{ab}$ and $C_{ab}$). We find it useful to define
\begin{align}
\langle a \rangle \equiv \int_0^1 du \, a(u), \,
\end{align}
and
\begin{align}
    [a](u) \equiv - \int_0^u dv a(v) + u a(u)\,.
\end{align}
For the replica-symmetric case, we take 
\begin{align}
a(u) = a, \quad\quad \tilde{a} = a + \sigma_a\,.
\end{align}
For the case of one-step replica symmetry breaking, we take 
\begin{align}
\tilde{a} & = a_1 + \sigma_a, \nonumber \\
    a(u) & = 
    \begin{cases}
    a_1 \,, \quad \mbox{$x < u < 1$} \\
    a_0 \,, \quad \mbox{$0 < u < x$}
    \end{cases}.
\end{align}
Then, we have $\langle a \rangle = a_1 -x (a_1-a_0)$, and
\begin{align}
    [a](u)  = 
    \begin{cases}
    x(a_1 - a_0)  \,, \quad &\mbox{$x < u < 1$} \\
    0 \,, \quad &\mbox{$0 < u < x$}
    \end{cases}.
\end{align}

We now present some useful identities for matrix operations in the limit $n \rightarrow 0$.

\subsection{Term-by-term sum}
\begin{equation}
    \frac{1}{n} \sum_{a,b=1}^n A_{ab} = \tilde{a} - \langle a \rangle. \label{ttsum}
\end{equation}

\subsection{Matrix product}

For
\begin{equation}
C_{ab} = \sum_{c=1}^n A_{ac} B_{cb}\,,
\end{equation}
we have \cite{MP91}
\begin{align}
\tilde{c} &= \tilde{a} \tilde{b} - \langle a b \rangle, \nonumber \\
c(u) &= (\tilde{b} - \langle b \rangle) a(u) + (\tilde{a} - \langle a \rangle) b(u) - \int_0^u dv (a(u) - a(v)) (b(u) - b(v)). \label{e1}
\end{align}

\subsection{Matrix inverse}

To determine $B = A^{-1}$, we set $C=1$ in the matrix product to obtain the relations
\begin{align}
1 &= \tilde{a} \tilde{b} - \langle a b \rangle, \nonumber \\
0 &= (\tilde{b} - \langle b \rangle) a(u) + (\tilde{a} - \langle a \rangle) b(u) - \int_0^u dv (a(u) - a(v)) (b(u) - b(v)). \label{ea1a}
\end{align}

With replica-symmetric matrices, we have
\begin{align}
\sigma_b &= \frac{1}{\sigma_a}, \nonumber \\
b &= - \frac{a}{\sigma_a^2}.
\end{align}

For the case of one-step replica symmetry breaking, we have
\begin{align}
    1 & = \tilde{a} \tilde{b} - a_0 b_0 x - a_1 b_1 (1-x), \nonumber \\
    0 & = (\tilde{b} - b_0 x - b_1 (1-x)) a_0 + (\tilde{a} - a_0 x - a_1 (1-x)) b_0,  \nonumber \\
    0 & = (\tilde{b} - b_0 x - b_1 (1-x)) a_1 + (\tilde{a} - a_0 x - a_1 (1-x)) b_1 - x (a_1 - a_0)(b_1-b_0). \label{product}
\end{align}
Solving these equations, we obtain
\begin{align}
\sigma_b &= \frac{1}{\sigma_a}, \nonumber \\
b_1 &=- \frac{a_1 \sigma_a + x (a_1-a_0)^2}{\sigma_a (\sigma_a + x (a_1-a_0))^2}, \nonumber \\
b_0 &= - \frac{a_0}{(\sigma_a + x (a_1-a_0))^2}.
\end{align}

\subsection{Term-by-term inverse sum}

If $A^{-1} = B$, then from (\ref{ttsum})
\begin{align}
    \frac{1}{n} \sum_{ab} A^{-1}_{ab} = \tilde{b} - \langle b \rangle.
\end{align}
From (\ref{ea1a}), we have
\begin{align}
    1 & = \tilde{a} \tilde{b} - \langle a b \rangle, \nonumber \\
    0 & = (\tilde{b} - \langle b \rangle) \langle a \rangle + (\tilde{a} - \langle a \rangle) \langle b \rangle - \langle ab \rangle + \langle a \rangle\langle b \rangle.
\end{align}
Taking the difference, we obtain
\begin{align}
    \frac{1}{n} \sum_{ab} A^{-1}_{ab} = \frac{1}{\tilde{a} - \langle a \rangle}.
\end{align}

\subsection{Term-by-term product}
For the product
\begin{equation}
D_{ab} = \sum_{c,d=1}^n A_{ac} B_{db}\,, \label{e2}
\end{equation}
we use the fact that $\sum_{c}A_{ac}$ is independent of the index $a$ for replica matrices. This yields a replica-symmetric matrix $D_{ab}$, which is of independent $a,b$ and using (\ref{ttsum}) we have
\begin{align}
\tilde{d} = d(u) = (\tilde{a} - \langle a \rangle)(\tilde{b} - \langle b \rangle).
\end{align}
For the replica-symmetric case, $\tilde{d}=d = \sigma_a \sigma_b$, and $\sigma_d = 0$.
With one-step replica symmetry breaking $\tilde{d} = d_0 = d_1 = (\sigma_a + x(a_1-a_0))(\sigma_b + x(b_1-b_0))$, and $\sigma_d = 0$.

\subsection{Trace log}

The trace log is needed in the computation of the free energy.
We have \cite{MP91}
\begin{align}
\frac{1}{n} \mbox{Tr} \ln A = \ln ( \tilde{a} - \langle a \rangle ) + \frac{a(0)}{\tilde{a} - \langle a \rangle} - \int_0^1 \frac{du}{u^2} \ln \frac{\tilde{a} - \langle a \rangle - [a](u)}{\tilde{a} - \langle a \rangle}\,. \label{trln}
\end{align}

In the replica-symmetric case, 
we have
\begin{align}
\frac{1}{n} \mbox{Tr} \ln A = \ln(\sigma_a) + \frac{a}{\sigma_a} \,.
\end{align}

With one-step replica symmetry breaking, we have 
\begin{align}
\frac{1}{n} \mbox{Tr} \ln A = \ln ( \sigma_a) + \frac{a_0}{\sigma_a + x (a_1-a_0)} -\frac{1}{x} \ln \frac{\sigma_a }{\sigma_a + x (a_1-a_0)}\,. \label{trln2}
\end{align}

Let us now consider $\mbox{Tr} \ln (A+B)$, where $B$ is a constant matrix, {\it i.e.\/}, $\tilde{b} = b(u) = b$, and $\sigma_b = 0$. Then, we have
\begin{align}
\frac{1}{n} \mbox{Tr} \ln (A + B) = \frac{1}{n} \mbox{Tr} \ln A + \frac{b}{\tilde{a} - \langle a \rangle}. \label{CSidentity}
\end{align}

\subsection{Cubic term}

According to (3.72) in Ref.~\onlinecite{HertzBook}, for the 
matrix $q_{ab}$ with $\tilde{q} = 0$,
\begin{align}
\frac{1}{n} \mbox{Tr} q^3 = \int_0^1 du \left[ u [q(u)]^3 + 3 q(u) \int_0^u dv [q(v)]^2 \right].
\end{align}

\bibliography{refs}

\end{document}